\documentclass[letterpaper,11pt]{article}
\usepackage{amsmath,amssymb,color,epsfig}
\usepackage{yfonts}
\usepackage{graphicx}
\usepackage{subfigure}
\usepackage{setspace}
\usepackage{physics}
\usepackage{amsfonts}
\usepackage{mathtools}
\usepackage{tikz-cd}
\usepackage{tikz}
\usetikzlibrary{decorations.markings}
\usepackage{framed}
\usetikzlibrary{shapes.geometric, arrows, positioning}
\usepackage{empheq}

\pdfoutput=1
\usepackage{jheppub}
\usepackage[T1]{fontenc}
\newcommand{\be}{\begin{equation}}
\newcommand{\ee}{\end{equation}}
\newcommand{\bea}{\setlength\arraycolsep{2pt} \begin{eqnarray}}
\newcommand{\eea}{\end{eqnarray}}
\newcommand{\nn}{\nonumber}

\def\0{{\sst{(0)}}}
\def\1{{\sst{(1)}}}
\def\2{{\sst{(2)}}}
\def\3{{\sst{(3)}}}
\def\4{{\sst{(4)}}}
\def\5{{\sst{(5)}}}
\def\6{{\sst{(6)}}}
\def\7{{\sst{(7)}}}
\def\8{{\sst{(8)}}}
\def\sst#1{{\scriptscriptstyle #1}}

\hypersetup{
    colorlinks=true,
    linkcolor=blue,
    citecolor=blue,
    urlcolor=blue
}

\title{Time-Ordering in the Dyson–Unruh Problem:\\ Accelerated Observers and Quantum Fields
}

\author[a]{Arash Azizi}

\affiliation[a]{{\it The Institute for Quantum Science and Engineering,
Texas A\&M University,\\ College Station, TX 77843, U.S.A.}}

\emailAdd{sazizi@tamu.edu}

\abstract{We develop a systematic framework for analyzing time-ordered perturbative expansions in quantum field theory (QFT) in curved spacetime, focusing on the interaction between a scalar field and multiple Unruh--DeWitt detectors undergoing uniform acceleration. Assuming the presence of a timelike Killing vector field, we adopt it as a global time parameter and express each detector’s proper time accordingly to ensure consistent implementation of time-ordering in the Dyson expansion. Employing light-cone coordinates and explicit parametrizations of detector worldlines, we construct and classify all relevant interaction terms at first, second, and third order in perturbation theory for a two-detector system. Our formalism resolves time-ordering ambiguities that arise due to differing accelerations and causal relations between detectors, enabling a clear and unambiguous identification of S-matrix contributions across multiple perturbative orders.}

\begin{document} 
\maketitle
\flushbottom

\section{Introduction}

Quantum field theory (QFT) in curved spacetime~\cite{
Unruh1976,
Hawking1975,
Fulling1973,
DEWITT1975, Wald:1975kc, Fulling:1989nb, BirrellDavies1982, Wald1994, Witten22QFT, Parker2009, Mukhanov2007, HollandsWald2015} provides an essential intermediate framework in the pursuit of a complete theory of quantum gravity. Among the conceptual challenges in quantum gravity, the treatment of time emerges as particularly significant. Accordingly, the notion of time requires careful consideration even in the semiclassical regime of QFT on curved backgrounds.

The standard approach treats the spacetime geometry as a classical entity governed by general relativity, while quantum fields propagate on this fixed background. At first glance, one might expect the familiar formalism of QFT in flat Minkowski spacetime to extend naturally to curved spacetimes. However, a fundamental distinction arises in the treatment of time. In quantum mechanics, time appears explicitly as an external parameter in the Schrödinger equation. In contrast, general relativity treats time and space as dynamical components of the spacetime manifold, with no preferred temporal direction.

This mismatch raises the question: how should one define time evolution in QFT on a curved background? A widely adopted strategy involves identifying a timelike Killing vector field—denoted $\partial_t$ when available—to define a global time coordinate $t$. The spacetime can then be foliated into a family of spacelike Cauchy hypersurfaces labeled by this coordinate. In such a setup, the Dyson series provides a natural framework for implementing time evolution perturbatively.

A particularly delicate issue arises when coupling quantum fields to localized observers or detectors. Since the seminal work of Unruh~\cite{Unruh1976}, it has become clear that particle creation and annihilation acquire operational meaning only relative to an observer. The Unruh--DeWitt detector model~\cite{Unruh1976, Einstein100, LoukoSatz2008} provides a widely used and physically transparent framework in this context. These detectors, modeled as two-level systems, interact locally with the field along their worldlines.

Second-order processes involving Unruh--DeWitt detectors have attracted increasing attention, particularly in investigations of entanglement harvesting~\cite{Valentini1991, Reznik2003, Reznik2005, FuentesSchuller2005, Salton:2014jaa,
Pozas2015,
Sachs2017,
Zhang2020},   and correlations between two detectors in the same and opposite Rindler wedges \cite{Svidzinsky21prr, Svidzinsky21prl}. These studies emphasize the subtle role of time ordering in non-inertial frames and curved backgrounds. 

In this work, we consider multiple Unruh--DeWitt detectors coupled to a real Klein–Gordon field in $(1+1)$-dimensional spacetime. We assume the detectors are arbitrarily massive and follow classical trajectories parameterized by proper times $\tau_i$. This assumption allows the interaction Hamiltonians to be sharply localized in spacetime and conveniently expressed in terms of light-cone coordinates $(u, v)$. A central technical challenge arises when computing the Dyson expansion at higher orders: namely, the need to consistently enforce time-ordering across detectors with distinct accelerations and causal structures.

We assume the existence of a global timelike coordinate $t$ and express each detector’s proper time $\tau_i$ as a function of $t$. This enables a consistent implementation of time ordering by comparing the corresponding light-cone coordinates of the detectors. To carry this out explicitly, we introduce four auxiliary functions—$f(z)$, $g(z)$, $h(z)$, and $k(z)$—which encode causal relationships between interactions along different worldlines. These functions translate global time ordering into well-defined integration limits over light-cone variables, even when detector accelerations differ.

With this formalism, we construct all second-order Dyson terms and analyze the complete third-order structure. Each third-order term is labeled as $XYZ_{ijk}$, where $\{X,Y,Z\} \in \{R,L\}$ indicates the field mode type (right- or left-traveling), and $\{i,j,k\} \in \{1,2\}$ specifies the detectors involved at each interaction time. A complete listing and detailed construction of all 64 third-order contributions is provided in Appendix~\ref{64}.

The structure of the paper is as follows. In Section~\ref{observers}, we present the theoretical setup and describe how observers couple to the quantum field. Section~\ref{Dyson.2} formulates the Dyson series at second order in perturbation theory. Section~\ref{higher} generalizes the construction to arbitrary Dyson orders. The full classification of third-order terms is summarized in structure, with explicit integrals deferred to Appendix~\ref{64}. We conclude with a summary and outlook in Section~\ref{conc}.

\section{Observers coupled to  quantum field} \label{observers}

\subsection{An observer coupled to quantum field}

The interaction between an observer and a quantum field is commonly modeled by a Hamiltonian of the form
\begin{align}
H(t) = \int_{\Sigma_{d-1}} d^{d-1}x \, \Phi(x^\mu) \, \Psi_D(x^\mu) \, \delta^{(d-1)}\bigl(\mathbf{x} - \mathbf{x}_D(\tau)\bigr),
\end{align}
where the integration is taken over a Cauchy hypersurface \(\Sigma_{d-1}\) at fixed coordinate time \(t\). The spacetime coordinates are denoted \(x^\mu = (\mathbf{x}, t)\), with \(\mathbf{x}\) representing the spatial components. The field \(\Phi(x^\mu)\) is a massless Klein-Gordon scalar field, and \(\Psi_D(x^\mu)\) represents the observer’s wave function.

A crucial point here is that the observer, assuming the simplest non-trivial model, i.e., a two-level atom, or Unruh--DeWitt detector, is not described by a second-quantized field theory, but rather treated within the framework of quantum mechanics. This treatment is justified by assuming the observer to be sufficiently massive so as to follow a sharply defined classical trajectory \(\mathbf{x}_D(\tau)\), parameterized by its proper time \(\tau\). In this semiclassical picture, the observer interacts locally with the field only along its worldline. The observer’s wave-function reads
\begin{align}
\Psi_D(\tau)= \sigma \, e^{- i \omega \tau } + \sigma^\dagger \, e^{i \omega \tau },
\end{align}
where \(\sigma\) denotes the lowering operator, which can be written explicitly as \(\sigma = \ket{g}\bra{e}\). Here, \(\ket{e}\) and \(\ket{g}\) represent the excited and ground states of the two-level  Unruh--DeWitt  detector, which serves as the observer in this framework. The parameter \(\tau\) denotes the observer's proper time, and \(\omega\) is the transition frequency between energy levels, such that \(\hbar \omega\) gives the energy difference between the two states.

The detector follows a classical worldline \(\mathbf{x}_D(\tau)\), and the trajectory is parametrized by \(\tau\), which may either correspond to proper time or another monotonic parameter along the curve. The delta function in the interaction Hamiltonian enforces strict locality by ensuring that the quantum field couples to the detector precisely at its location on the Cauchy hypersurface.

Evaluating the spatial integral in the interaction Hamiltonian yields
\begin{align}
H(t) = \Phi\bigl(\mathbf{x}_D(\tau), t\bigr) \, \Psi_D\bigl(\mathbf{x}_D(\tau), t\bigr),
\end{align}
where the field and detector wavefunction are evaluated at the observer’s location on the hypersurface labeled by coordinate time \(t\). If the coordinate time \(t\) can be expressed as a monotonic function of the observer’s proper time \(\tau\), then the inverse function \(\tau(t)\) exists and allows the Hamiltonian to be rewritten entirely in terms of \(t\):
\begin{align}
H(t) = \Phi\bigl(\mathbf{x}_D(t), t\bigr) \, \Psi_D\bigl(\mathbf{x}_D(t), t\bigr),
\end{align}
providing a consistent and covariant formulation of the observer-field interaction in a curved background, with all quantities expressed in terms of the global time coordinate.

To make the discussion more explicit, we consider the simplest possible case: a real scalar Klein-Gordon field in \(1+1\) dimensions. The field can be decomposed into its right- and left-moving components in light-cone coordinates as
\begin{align}
\Phi(x,t) 
=&\, \Phi_{\text{RTW}}(u) + \Phi_{\text{LTW}}(v) \nn\\
=&\, \int_0^{\infty} \frac{d\nu}{\sqrt{4\pi \nu}} \left( e^{-i \nu u} a_\nu + e^{i \nu u} a^\dagger_\nu \right)
+ \int_0^{\infty} \frac{d\nu}{\sqrt{4\pi \nu}} \left( e^{-i \nu v} b_\nu + e^{i \nu v} b^\dagger_\nu \right).
\end{align}
The light-cone coordinates \(u\) and \(v\) are defined by \(u = t - x\) and \(v = t + x\), where we have set \(c = 1\). Moreover, \(a_\nu\) \((b_\nu)\) and \(a_\nu^\dagger\) \((b_\nu^\dagger)\) are the annihilation and creation operators corresponding to right-moving (left-moving) modes, respectively, and satisfy the standard commutation relations.

\subsection{Two observers coupled to a quantum field}

\begin{figure}[ht]
\centering
\includegraphics[width=1\textwidth]{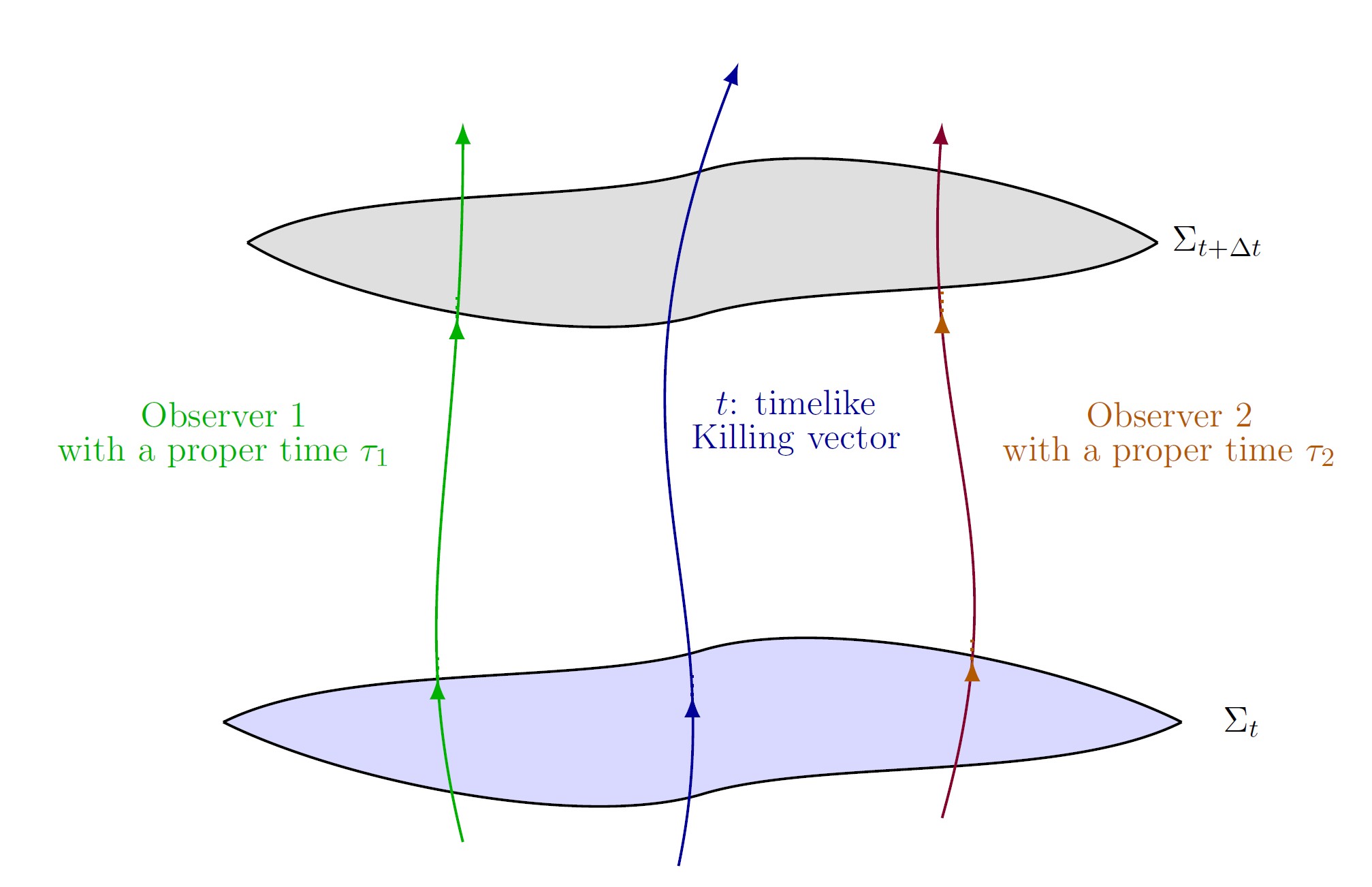}
\caption{A schematic illustration of two observers coupled to a quantum field in a spacetime admitting a timelike Killing vector $\partial_t$. The two hypersurfaces $\Sigma_t$ and $\Sigma_{t+\Delta t}$ represent constant-time slices associated with this coordinate. Each observer follows a distinct worldline with respective proper times $\tau_1$ and $\tau_2$, both of which can be expressed in terms of the global coordinate time $t$. This setup allows for covariant formulation of time evolution and facilitates time-ordering analysis in the Dyson series expansion.}
\label{fig:observer-setup}
\end{figure}

As before, we assume that each observer interacts locally with the quantum field and is sufficiently massive to be treated as a point particle following a well-defined classical trajectory. The total Hamiltonian governing the interaction between the field and two such observers can be expressed as
\begin{align}
H(t) 
=& \int_{\Sigma_{d-1}} d^{d-1}x \, \Phi(x^\mu) \, \Psi_D(x^\mu) 
\left( 
\delta^{(d-1)}\big(\mathbf{x} - \mathbf{x}_{D_1}(\tau_1)\big) 
+ 
\delta^{(d-1)}\big(\mathbf{x} - \mathbf{x}_{D_2}(\tau_2)\big) 
\right) \nn\\
=&\, \Phi\big(\mathbf{x}_{D_1}(\tau_1),t\big) \, \Psi_{D_1}\big(\mathbf{x}_{D_1}(\tau_1),t\big) 
+ 
\Phi\big(\mathbf{x}_{D_2}(\tau_2),t\big) \, \Psi_{D_2}\big(\mathbf{x}_{D_2}(\tau_2),t\big) \nn\\
=&\, H_1(t) + H_2(t),
\end{align}
where \(H_1(t)\) and \(H_2(t)\) denote the individual interaction Hamiltonians of the two observers (see the figure \ref{fig:observer-setup}).

The crucial observation is that the observers are synchronized in terms of the global Killing time coordinate \(t\), such that
\begin{align}
t(\tau_1) = t(\tau_2) = t.
\end{align}
This allows all quantities to be expressed consistently in terms of the coordinate time \(t\), yielding a compact and covariant form for the total Hamiltonian:
\begin{align}
H(t) 
= \Phi\big(\mathbf{x}_{D_1}(t),t\big) \, \Psi_{D_1}\big(\mathbf{x}_{D_1}(t),t\big) 
+ 
\Phi\big(\mathbf{x}_{D_2}(t),t\big) \, \Psi_{D_2}\big(\mathbf{x}_{D_2}(t),t\big).
\end{align}

\subsection{The case of uniform acceleration}
One of the most well-studied scenarios in quantum field theory on curved (or flat)  backgrounds is the case of a particle undergoing uniform acceleration. While Einstein and Rosen remarked in 1935 that the problem of uniform acceleration was already well-known at the time \cite{ER35}, it was Rindler who later formalized and publicized the coordinate system adapted to uniformly accelerated observers \cite{Rindler66}.

The trajectory of a uniformly accelerated particle in Minkowski spacetime, with proper acceleration \(a\), can be described by the parametric equations:
\begin{align}
x = \frac{1}{a}\,  \cosh(a \tau)\,, \qquad
t = \frac{1}{a}\,\sinh(a \tau)\,,
\end{align}
where \(\tau\) is the proper time  along the trajectory. A useful identity within the right Rindler wedge—where \(u < 0\) and \(v > 0\)—is given by:
\begin{align}
-au = e^{-a \tau}, \qquad av = e^{a \tau}.
\end{align}
These relations express the light-cone coordinates \(u = t - x\) and \(v = t + x\) in terms of the observer’s proper time \(\tau\), and will prove particularly useful in formulating field interactions and detector responses for uniformly accelerated observers.

\subsection{Resolving the Jacobian}

A subtle complication arises when changing variables from the observer’s proper time \(\tau\) to the global Killing time coordinate \(t\), due to the presence of a Jacobian factor in the integral measure. However, this difficulty can be circumvented by introducing a slight modification to the form of the interaction Hamiltonian. Specifically, consider the following expression:
\begin{align}
H(t) = \frac{\partial}{\partial t} \Phi(\mathbf{x}_D(t), t) \, \Psi_D(\mathbf{x}_D(t), t),
\end{align}
where the time derivative acts explicitly on the field operator. In this formulation, the Jacobian associated with the change of variables is effectively absorbed, thanks to the identity:
\begin{align}
dt \, \partial_t = d\tau \, \partial_\tau.
\end{align}

This identity ensures that time evolution is preserved consistently under reparametrization. Moreover, there is a physical motivation for this form of the Hamiltonian: it closely resembles the dipole interaction in electrodynamics, where the interaction Hamiltonian is proportional to the electric field—i.e., the time derivative of the vector potential. By analogy, the scalar field \(\Phi\) here plays a role similar to the vector potential, and the time derivative \(\partial_t \Phi\) enters as the physically relevant observable in the interaction. One may find this Hamiltonian in many papers, for instance \cite{Svidzinsky21prr}.

\section{Time evolution: Dyson's series} \label{Dyson.2}

\subsection{First order}

In the case of uniform acceleration, Minkowski time \(t\) can be expressed as a monotonic function of the proper time \(\tau\). Since both parametrizations span the entire real line, one can consistently use the global time coordinate \(t\) as a common parameter along the trajectories of all observers. As a result, \(t\) may serve as the evolution parameter in the Dyson series expansion. The interaction Hamiltonian for the \(k^{\text{th}}\) observer is given by
\begin{align}
H_k(t) = g \left( 
\frac{\partial}{\partial u_k} \Phi_{\text{RTW}}(u_k(t)) 
+ 
\frac{\partial}{\partial v_k} \Phi_{\text{LTW}}(v_k(t)) 
\right) 
\left( 
\sigma \, e^{-i\omega \tau_k(t)} 
+ 
\sigma^\dagger \, e^{i\omega \tau_k(t)} 
\right),
\end{align}
where \(u_k(t)\) and \(v_k(t)\) are the light-cone coordinates along the \(k^{\text{th}}\) trajectory, expressed in terms of the global time coordinate \(t\).

Summing over all observers, the first-order Dyson integral becomes:
\begin{align}
\int dt \sum_{k=1}^{m} H_k(t) 
=&\, g \sum_{k=1}^{m} \int_{{u_k}_i}^{{u_k}_f} du_k \, 
\frac{\partial}{\partial u_k} \Phi_{\text{RTW}}(u_k) 
\left( 
\sigma \, e^{-i\omega \tau_k(u_k)} 
+ 
\sigma^\dagger \, e^{i\omega \tau_k(u_k)} 
\right) \nn\\
&+\, g \sum_{k=1}^{m} \int_{{v_k}_i}^{{v_k}_f} dv_k \, 
\frac{\partial}{\partial v_k} \Phi_{\text{LTW}}(v_k) 
\left( 
\sigma \, e^{-i\omega \tau_k(v_k)} 
+ 
\sigma^\dagger \, e^{i\omega \tau_k(v_k)} 
\right).
\end{align}

Here, no complications arise in changing variables, since the differential operators satisfy the equivalence
\[
dt \, \partial_t = du \, \partial_u = dv \, \partial_v,
\]
and the integration limits for $t$ extend over the full real line, from \(-\infty\) to \(+\infty\), ensuring a well-defined formulation at first order in perturbation theory.

\subsubsection{Two detectors in the same wedge} 

Assume both detectors are located in the right Rindler wedge, with their trajectories parametrized by
\[
-a_1 u_1 = e^{-a_1 \tau_1}, \qquad
a_1 v_1 = e^{a_1 \tau_1}, \qquad
-a_2 u_2 = e^{-a_2 \tau_2}, \qquad
a_2 v_2 = e^{a_2 \tau_2}.
\]
Then, the first-order contribution to Dyson's series becomes
\begin{align}
\int_{-\infty}^{+\infty} dt \, H(t) 
=&\, g \int_{-\infty}^{0} du \, \frac{\partial}{\partial u} \Phi_{\text{RTW}}(u) 
\left( \sigma_1 (-a_1 u)^{i \frac{\omega}{a_1}} + \sigma_1^\dagger (-a_1 u)^{-i \frac{\omega}{a_1}} \right) \nonumber \\
&+\, g \int_{0}^{\infty} dv \, \frac{\partial}{\partial v} \Phi_{\text{LTW}}(v) 
\left( \sigma_1 (a_1 v)^{-i \frac{\omega}{a_1}} + \sigma_1^\dagger (a_1 v)^{i \frac{\omega}{a_1}} \right) \nonumber \\
&+\, g \int_{-\infty}^{0} du \, \frac{\partial}{\partial u} \Phi_{\text{RTW}}(u) 
\left( \sigma_2 (-a_2 u)^{i \frac{\omega}{a_2}} + \sigma_2^\dagger (-a_2 u)^{-i \frac{\omega}{a_2}} \right) \nonumber \\
&+\, g \int_{0}^{\infty} dv \, \frac{\partial}{\partial v} \Phi_{\text{LTW}}(v) 
\left( \sigma_2 (a_2 v)^{-i \frac{\omega}{a_2}} + \sigma_2^\dagger (a_2 v)^{i \frac{\omega}{a_2}} \right). \label{1st.order.same}
\end{align}

\subsubsection{Two detectors in opposite wedges} 

Now consider the case where the first detector lies in the right Rindler wedge and the second in the left wedge. Their trajectories are parameterized by
\[
-a_1 u_1 = e^{-a_1 \tau_1}, \qquad
a_1 v_1 = e^{a_1 \tau_1}, \qquad
a_2 u_2 = e^{a_2 \tau_2}, \qquad
-a_2 v_2 = e^{-a_2 \tau_2}.
\]
Then, the first-order contribution to Dyson's series takes the form:
\begin{align}
\int_{-\infty}^{+\infty} dt \, H(t) 
=&\, g \int_{-\infty}^{0} du \, \frac{\partial}{\partial u} \Phi_{\text{RTW}}(u) 
\left( \sigma_1 (-a_1 u)^{i \frac{\omega}{a_1}} + \sigma_1^\dagger (-a_1 u)^{-i \frac{\omega}{a_1}} \right) \nonumber \\
&+\, g \int_{0}^{\infty} dv \, \frac{\partial}{\partial v} \Phi_{\text{LTW}}(v) 
\left( \sigma_1 (a_1 v)^{-i \frac{\omega}{a_1}} + \sigma_1^\dagger (a_1 v)^{i \frac{\omega}{a_1}} \right) \nonumber \\
&+\, g \int_{0}^{\infty} du \, \frac{\partial}{\partial u} \Phi_{\text{RTW}}(u) 
\left( \sigma_2 (a_2 u)^{-i \frac{\omega}{a_2}} + \sigma_2^\dagger (a_2 u)^{i \frac{\omega}{a_2}} \right) \nonumber \\
&+\, g \int_{-\infty}^{0} dv \, \frac{\partial}{\partial v} \Phi_{\text{LTW}}(v) 
\left( \sigma_2 (-a_2 v)^{i \frac{\omega}{a_2}} + \sigma_2^\dagger (-a_2 v)^{-i \frac{\omega}{a_2}} \right). \label{1st.order.opp}
\end{align}

\subsection{Second order} 
The second-order term in the Dyson expansion introduces significant complexity due to the requirement of time ordering. In contrast to the first-order case—where the integration runs over the entire real line without temporal constraints—second-order contributions must carefully respect the causal structure of interactions, especially in the presence of multiple detectors.

In the simplest nontrivial scenario involving two detectors, the Hamiltonian at each time consists of eight distinct terms. This count arises from two detectors, each with two internal operators, coupled to two field modes: one right-traveling wave (RTW) and one left-traveling wave (LTW). Consequently, the second-order expansion contains a total of \(8 \times 8 = 64\) terms.

The second-order Dyson term takes the form
\begin{align}
\int_{-\infty}^{+\infty} dt \, H(t) \int_{-\infty}^{t} dt' \, H(t'),
\end{align}
so that the full second-order evolution operator is given by
\begin{align}
S_{2,2} 
=& \left( -\frac{i}{\hbar} \right)^2 \int_{-\infty}^{\infty} dt \int_{-\infty}^{t} dt' \, H(t) H(t') \\
=& \left( -\frac{i}{\hbar} \right)^2 \int_{-\infty}^{\infty} dt \, \big(H_1(t) + H_2(t)\big) \int_{-\infty}^{t} dt' \, \big(H_1(t') + H_2(t')\big) \nn\\
=& \left( -\frac{i}{\hbar} \right)^2 \Bigg\{ \int_{-\infty}^{\infty} dt \int_{-\infty}^{t} dt' \Big( 
H_1(t) H_1(t') 
+ H_1(t) H_2(t') 
+ H_2(t) H_1(t') 
+ H_2(t) H_2(t') 
\Big) \Bigg\}. \nn
\end{align}
Here, we have introduced the notation \(S_{2,2}\) to denote the second-order Dyson contribution involving two detectors. Substituting the explicit form of the Hamiltonians derived earlier, we obtain
\begin{align}
S_{2,2} 
=& -\frac{1}{\hbar^2}\, \int_{-\infty}^{+\infty} dt \, 
\Bigg\{ 
\frac{\partial}{\partial t} \Phi\big(x_1^{\mu}(t)\big) 
\left( \sigma_1 e^{-i \omega \tau_1(t)} + \sigma_1^\dagger e^{i \omega \tau_1(t)} \right) \nonumber \\
&\phantom{-\frac{1}{\hbar^2}\, \int_{-\infty}^{+\infty} dt \, 
\Bigg\{ } 
+ \frac{\partial}{\partial t} \Phi\big(x_2^{\mu}(t)\big) 
\left( \sigma_2 e^{-i \omega \tau_2(t)} + \sigma_2^\dagger e^{i \omega \tau_2(t)} \right) 
\Bigg\} \nonumber \\
&\quad \quad \times \int_{-\infty}^{t} dt' \, 
\Bigg\{ 
\frac{\partial}{\partial t'} \Phi\big(x_1^{\mu}(t')\big) 
\left( \sigma_1 e^{-i \omega \tau_1(t')} + \sigma_1^\dagger e^{i \omega \tau_1(t')} \right) \nonumber \\
&\phantom{-\frac{1}{\hbar^2}\, \int_{-\infty}^{+\infty} dt \, 
\Bigg\{ } 
+ \frac{\partial}{\partial t'} \Phi\big(x_2^{\mu}(t')\big) 
\left( \sigma_2 e^{-i \omega \tau_2(t')} + \sigma_2^\dagger e^{i \omega \tau_2(t')} \right) 
\Bigg\}. \label{2nd.order}
\end{align}
Since each field \(\Phi\) consists of both right- and left-moving components, the total number of operator combinations in \(S_{2,2}\) indeed amounts to 64. One may divide these 64 terms into 4 distinct types based on the right- or left- traveling wave of the field modes. Namely, 
\begin{align}
S_{2,2} = S_{2,2}^{\text{RTW--RTW}}
+S_{2,2}^{\text{RTW--LTW}}
+S_{2,2}^{\text{LTW--RTW}}
+S_{2,2}^{\text{LTW--LTW}}.
\end{align}
where each of the above expressions contains 16 terms. In the following subsections, we systematically analyze the contributions from these various mode combinations.

\subsubsection{Same wedge detectors: right-traveling wave–right-traveling wave fields}
We now consider the case in which both detectors are located in the right Rindler wedge and couple exclusively to the right-traveling components of the field (RTW). In this configuration, all second-order contributions involve products of RTW modes, denoted by \(RR_{ij}\), where \(i, j = 1, 2\) label the detectors.

A subtlety arises when converting the time-ordering condition \(t' < t\) into light-cone variables \(u_1\) and \(u_2\). To determine the upper limit of the second integral, one must solve the condition
\begin{align}
t(u_1) = t(u_2),
\end{align}
which, using the Rindler parametrization for uniformly accelerated trajectories, yields the equation
\begin{align}
\frac{1}{2} \left( u_1 - \frac{1}{a_1^2 u_1} \right) = 
\frac{1}{2} \left( u_2 - \frac{1}{a_2^2 u_2} \right). \label{same.time}
\end{align}
Solving (\ref{same.time}) for \(u_2\) in terms of \(u_1\) gives
\begin{align}
u_2 = \frac{1}{2} \left( u_1 - \frac{1}{a_1^2 u_1} \right) 
\pm \frac{1}{2} \sqrt{ \left( u_1 - \frac{1}{a_1^2 u_1} \right)^2 + \frac{4}{a_2^2} }. \label{2 in terms of 1}
\end{align}
It is convenient to define two functions:
\begin{align}
f(z) &= \frac{1}{2} \left( z - \frac{1}{a_1^2 z} \right) 
+ \sqrt{ \frac{1}{4} \left( z - \frac{1}{a_1^2 z} \right)^2 + \frac{1}{a_2^2} }, \nn\\
g(z) &= \frac{1}{2} \left( z - \frac{1}{a_1^2 z} \right) 
- \sqrt{ \frac{1}{4} \left( z - \frac{1}{a_1^2 z} \right)^2 + \frac{1}{a_2^2} }, \label{f.g}
\end{align}
where \(f(z)\) and \(g(z)\) represent the positive and negative branches of the solution to the quadratic equation (\ref{2 in terms of 1}), respectively.


Moreover, solving (\ref{same.time}) for \(u_1\) in terms of \(u_2\) gives
\begin{align}
u_1 = \frac{1}{2} \left( u_2 - \frac{1}{a_2^2 u_2} \right) 
\pm \sqrt{ \frac{1}{4}  \left( u_2 - \frac{1}{a_2^2 u_2} \right)^2 + \frac{1}{a_1^2} }. \label{1 in terms of 2}
\end{align}
Again, let's introduce the following functions:
\begin{align}
h(z) &= \frac{1}{2} \left( z - \frac{1}{a_2^2 z} \right) 
+ \sqrt{ \frac{1}{4} \left( z - \frac{1}{a_2^2 z} \right)^2 + \frac{1}{a_1^2} }, \nn\\
k(z) &= \frac{1}{2} \left( z - \frac{1}{a_2^2 z} \right) 
- \sqrt{ \frac{1}{4} \left( z - \frac{1}{a_2^2 z} \right)^2 + \frac{1}{a_1^2} }, \label{h.k}
\end{align}
where \(h(z)\) and \(k(z)\) represent the positive and negative branches of the inverse relation to (\ref{1 in terms of 2}), corresponding to expressing \(u_1\) in terms of \(u_2\).

The full second-order RTW–RTW contribution is expressed as
\begin{align}
S_{2,2}^{\text{RTW--RTW}} = RR_{11} + RR_{12} + RR_{21} + RR_{22},
\end{align}
where each term \(RR_{ij}\) corresponds to a second-order interaction involving the \(i^{\text{th}}\) detector at time \(t\) and the \(j^{\text{th}}\) detector at earlier time \(t'\).

The diagonal terms \(RR_{ii}\), which involve only a single detector with fixed acceleration \(a_i\), present no complication. In the case where \(a_1 = a_2 = a\), the solution to (\ref{2 in terms of 1}) simplifies to \(u_1 = u_2\), and time-ordering can be directly implemented:
\begin{align}
RR_{ii} 
=& \left( -\frac{ig}{\hbar} \right)^2 
\int_{-\infty}^{0} du \, \frac{\partial}{\partial u} 
\Phi_{\text{RTW}}(u)
\left( \sigma_i (-a_i u)^{\frac{i \omega_i}{a_i}} + \sigma_i^\dagger (-a_i u)^{-\frac{i \omega_i}{a_i}} \right) \nonumber \\
& \phantom{\left( -\frac{ig}{\hbar} \right)^2}
\times \int_{-\infty}^{u} du' \, \frac{\partial}{\partial u'} 
\Phi_{\text{RTW}}(u')
\left( \sigma_i (-a_i u')^{\frac{i \omega_i}{a_i}} + \sigma_i^\dagger (-a_i u')^{-\frac{i \omega_i}{a_i}} \right). \label{RRii.same}
\end{align}
where $i=1,2$.

In contrast, the cross-term \(RR_{12}\) involves detectors with different accelerations, \(a_1 \neq a_2\), which introduces a nontrivial time-ordering boundary in the second integral. The correct upper limit in this case is the negative root \(g(u)\) from (\ref{f.g}), since \(u' < 0\):
\begin{align}
RR_{12} 
=& \left( -\frac{ig}{\hbar} \right)^2 
\int_{-\infty}^{0} du \, \frac{\partial}{\partial u} 
\Phi_{\text{RTW}}(u)
\left( \sigma_1 (-a_1 u)^{\frac{i \omega_1}{a_1}} + \sigma_1^\dagger (-a_1 u)^{-\frac{i \omega_1}{a_1}} \right) \nonumber \\
& \phantom{\left( -\frac{ig}{\hbar} \right)^2}
\times \int_{-\infty}^{g(u)} du' \, \frac{\partial}{\partial u'} 
\Phi_{\text{RTW}}(u')
\left( \sigma_2 (-a_2 u')^{\frac{i \omega_2}{a_2}} + \sigma_2^\dagger (-a_2 u')^{-\frac{i \omega_2}{a_2}} \right). \label{RR12.same}
\end{align}
where we have used $g(u)$ of (\ref{f.g}) since $u'<0$. Finally \(RR_{21}\) reads
\begin{align}
RR_{21} 
=& \left( -\frac{ig}{\hbar} \right)^2 
\int_{-\infty}^{0} du \, \frac{\partial}{\partial u} 
\Phi_{\text{RTW}}(u)
\left( \sigma_2 (-a_2 u)^{\frac{i \omega_2}{a_2}} + \sigma_2^\dagger (-a_2 u)^{-\frac{i \omega_2}{a_2}} \right) \nonumber \\
& \phantom{\left( -\frac{ig}{\hbar} \right)^2}
\times \int_{-\infty}^{k(u)} du' \, \frac{\partial}{\partial u'} 
\Phi_{\text{RTW}}(u')
\left( \sigma_1 (-a_1 u')^{\frac{i \omega_1}{a_1}} + \sigma_1^\dagger (-a_1 u')^{-\frac{i \omega_1}{a_1}} \right). \label{RR21.same}
\end{align}
where we have used $k(u)$ of (\ref{h.k}) since $u'<0$.

\subsubsection{Same wedge detectors: right-traveling wave–left-traveling wave fields}

In this subsection, we consider the second-order contribution to the Dyson series in which both detectors are located in the right Rindler wedge, but the interaction involves one right-traveling (RTW) and one left-traveling (LTW) component of the field. This configuration is denoted as RTW–LTW, and the corresponding contribution to the second-order S-matrix takes the form
\begin{align}
S_{2,2}^{\text{RTW--LTW}} = RL_{11} + RL_{12} + RL_{21} + RL_{22},
\end{align}
where \(RL_{ij}\) represents the contribution from the \(i^{\text{th}}\) detector at time \(t\) and the \(j^{\text{th}}\) detector at earlier time \(t'\).

To determine the integration limits consistent with time ordering, we equate the coordinate times associated with the RTW and LTW null coordinates:
\begin{align}
t(u) = t(v),
\end{align}
which leads to the condition
\begin{align}
\frac{1}{2} \left( u - \frac{1}{a_1^2 u} \right) = \frac{1}{2} \left( v - \frac{1}{a_2^2 v} \right).
\end{align}
Solving for \(v\) in terms of \(u\) yields
\begin{align}
v = \frac{1}{2} \left( u - \frac{1}{a_1^2 u} \right) 
\pm \sqrt{ \frac{1}{4} \left( u - \frac{1}{a_1^2 u} \right)^2 + \frac{1}{a_2^2} }. \label{2 in terms of 1.vu}
\end{align}
This is exactly the equation (\ref{2 in terms of 1}). We denote the positive and negative roots as \(f(u)\) and \(g(u)\), respectively, as defined earlier in (\ref{f.g}).

Moreover, for the following equation
\begin{align}
\frac{1}{2} \left( u - \frac{1}{a_2^2 u} \right) = \frac{1}{2} \left( v - \frac{1}{a_1^2 v} \right).
\end{align}
Solving for \(v\) in terms of \(u\) yields
\begin{align}
v = \frac{1}{2} \left( u - \frac{1}{a_2^2 u} \right) 
\pm \sqrt{ \frac{1}{4} \left( u - \frac{1}{a_2^2 u} \right)^2 + \frac{1}{a_1^2} }. \label{1 in terms of 2.vu}
\end{align}
This is exactly the equation (\ref{1 in terms of 2}). We denote the positive and negative roots as \(h(u)\) and \(k(u)\), respectively, as defined earlier in (\ref{h.k}).

In the special case of \(RL_{11}\), where both interactions involve the same detector with acceleration \(a\), the solutions simplify to
\begin{align}
v = \frac{1}{2} \left( u - \frac{1}{a^2 u} \right) 
\pm \frac{1}{2} \left( u + \frac{1}{a^2 u} \right)
\quad \Rightarrow \quad
f(u) = -\frac{1}{a^2 u}, \qquad g(u) = u.
\end{align}
Since both detectors are located in the right Rindler wedge, the acceptable upper limit corresponds to the positive solution \(f(u)\). Consequently, the term \(RL_{11}\), involving a RTW interaction at time \(t\) and an LTW interaction at earlier time \(t'\), becomes:
\begin{align}
RL_{11} 
=& \left( -\frac{ig}{\hbar} \right)^2 
\int_{-\infty}^{0} du \, \frac{\partial}{\partial u} 
\Phi_{\text{RTW}}(u)
\left( \sigma_1 (-a_1 u)^{\frac{i \omega_1}{a_1}} + \sigma_1^\dagger (-a_1 u)^{-\frac{i \omega_1}{a_1}} \right) \nonumber \\
& \phantom{\left( -\frac{ig}{\hbar} \right)^2}
\times \int_{0}^{-\frac{1}{a_1^2 u}} dv \, \frac{\partial}{\partial v} 
\Phi_{\text{LTW}}(v)
\left( \sigma_1 (a_1 v)^{-\frac{i \omega_1}{a_1}} + \sigma_1^\dagger (a_1 v)^{\frac{i \omega_1}{a_1}} \right). \label{RL11.same}
\end{align}
The mixed-detector contribution \(RL_{12}\), where the first detector acts at time \(t\) and the second at earlier time \(t'\), reads:
\begin{align}
RL_{12} 
=& \left( -\frac{ig}{\hbar} \right)^2 
\int_{-\infty}^{0} du \, \frac{\partial}{\partial u} 
\Phi_{\text{RTW}}(u)
\left( \sigma_1 (-a_1 u)^{\frac{i \omega_1}{a_1}} + \sigma_1^\dagger (-a_1 u)^{-\frac{i \omega_1}{a_1}} \right) \nonumber \\
&\phantom{\left( -\frac{ig}{\hbar} \right)^2}
\times \int_{0}^{f(u)} dv \, \frac{\partial}{\partial v} 
\Phi_{\text{LTW}}(v)
\left( \sigma_2 (a_2 v)^{-\frac{i \omega_2}{a_2}} + \sigma_2^\dagger (a_2 v)^{\frac{i \omega_2}{a_2}} \right), \label{RL12.same}
\end{align}
where the positive root \(f(u)\) from (\ref{2 in terms of 1.vu}) is selected to ensure that \(v > 0\), consistent with the domain of the detector in the right Rindler wedge. The mixed-detector contribution \(RL_{21}\) reads:
\begin{align}
RL_{21} 
=& \left( -\frac{ig}{\hbar} \right)^2 
\int_{-\infty}^{0} du \, \frac{\partial}{\partial u} 
\Phi_{\text{RTW}}(u)
\left( \sigma_2 (-a_2 u)^{\frac{i \omega_2}{a_2}} + \sigma_2^\dagger (-a_2 u)^{-\frac{i \omega_2}{a_2}} \right) \nonumber \\
&\phantom{\left( -\frac{ig}{\hbar} \right)^2}
\times \int_{0}^{h(u)} dv \, \frac{\partial}{\partial v} 
\Phi_{\text{LTW}}(v)
\left( \sigma_1 (a_1 v)^{-\frac{i \omega_1}{a_1}} + \sigma_1^\dagger (a_1 v)^{\frac{i \omega_1}{a_1}} \right), \label{RL21.same}
\end{align}
where the positive root \(h(u)\) from (\ref{1 in terms of 2}) is selected to ensure that \(v > 0\). Finally $RL_{22}$ reads
\begin{align}
RL_{22} 
=& \left( -\frac{ig}{\hbar} \right)^2 
\int_{-\infty}^{0} du \, \frac{\partial}{\partial u} 
\Phi_{\text{RTW}}(u)
\left( \sigma_2 (-a_2 u)^{\frac{i \omega_2}{a_2}} + \sigma_2^\dagger (-a_2 u)^{-\frac{i \omega_2}{a_2}} \right) \nonumber \\
& \phantom{\left( -\frac{ig}{\hbar} \right)^2}
\times \int_{0}^{-\frac{1}{a_2^2 u}} dv \, \frac{\partial}{\partial v} 
\Phi_{\text{LTW}}(v)
\left( \sigma_2 (a_2 v)^{-\frac{i \omega_2}{a_2}} + \sigma_2^\dagger (a_2 v)^{\frac{i \omega_2}{a_2}} \right). \label{RL22.same}
\end{align}

\subsubsection{Opposite wedge detectors: right-traveling wave–right-traveling wave fields}
We now consider the configuration in which the first detector is located in the right Rindler wedge and the second in the left wedge. In this subsection, we restrict the interaction to involve only the right-traveling wave (RTW) components of the field.

The term \(RR_{11}\) remains unchanged from the analysis in (\ref{RRii.same}), as it involves only the first detector in the right wedge. However, \(RR_{22}\) changes as follows
\begin{align}
RR_{22} 
=& \left( -\frac{ig}{\hbar} \right)^2 
\int_{0}^{\infty} du \, \frac{\partial}{\partial u} 
\Phi_{\text{RTW}}(u)
\left( \sigma_2 (a_2 u)^{-\frac{i \omega_2}{a_2}} + \sigma_2^\dagger (a_2 u)^{\frac{i \omega_2}{a_2}} \right) \nonumber \\
& \phantom{\left( -\frac{ig}{\hbar} \right)^2}
\times \int_{0}^{u} du' \, \frac{\partial}{\partial u'} 
\Phi_{\text{RTW}}(u')
\left( \sigma_2 (a_2 u')^{-\frac{i \omega_2}{a_2}} 
+ \sigma_2^\dagger (a_2 u')^{\frac{i \omega_2}{a_2}} \right). \label{RR22.op}
\end{align}

The cross-term \(RR_{12}\), in which the second detector lies in the left wedge, takes a different form due to the domain and parametrization of the second detector. The contribution \(RR_{12}\) is given by:
\begin{align}
RR_{12} 
=& \left( -\frac{ig}{\hbar} \right)^2 
\int_{-\infty}^{0} du \, \frac{\partial}{\partial u} 
\Phi_{\text{RTW}}(u)
\left( \sigma_1 (-a_1 u)^{\frac{i \omega_1}{a_1}} + \sigma_1^\dagger (-a_1 u)^{-\frac{i \omega_1}{a_1}} \right) \nonumber \\
&\phantom{\left( -\frac{ig}{\hbar} \right)^2}
\times \int_{0}^{f(u)} du' \, \frac{\partial}{\partial u'} 
\Phi_{\text{RTW}}(u')
\left( \sigma_2 (a_2 u')^{-\frac{i \omega_2}{a_2}} + \sigma_2^\dagger (a_2 u')^{\frac{i \omega_2}{a_2}} \right). \label{RR12.op}
\end{align}
Here \(u < 0\) denotes the coordinate of the first detector in the right wedge, while \(u' > 0\) corresponds to the left wedge trajectory of the second detector. To ensure proper time ordering, we must use the positive root of the matching condition in (\ref{2 in terms of 1})—that is, \(f(u)\) from (\ref{f.g})—as the upper limit of the second integral. This guarantees that the integration remains within the physical domain of the left wedge detector.

Finally, the term \(RR_{21}\), where the first detector lies on the right and the second detector lies on the left wedges, reads
\begin{align}
RR_{21} 
=& \left( -\frac{ig}{\hbar} \right)^2 
\int_{0}^{\infty} du \, \frac{\partial}{\partial u} 
\Phi_{\text{RTW}}(u)
\left( \sigma_2 (a_2 u)^{-\frac{i \omega_2}{a_2}} + \sigma_2^\dagger (a_2 u)^{\frac{i \omega_2}{a_2}} \right) \nonumber \\
&\phantom{\left( -\frac{ig}{\hbar} \right)^2}
\times \int_{-\infty}^{k(u)} du' \, \frac{\partial}{\partial u'} 
\Phi_{\text{RTW}}(u')
\left( \sigma_1 (-a_1 u')^{\frac{i \omega_1}{a_1}} 
+ \sigma_1^\dagger (-a_1 u')^{-\frac{i \omega_1}{a_1}} \right), \label{RR21.op}
\end{align}
where we have used $k(u)$ of (\ref{h.k}) since $u'<0$.

\subsubsection{Opposite wedge detectors: right-traveling wave–left-traveling wave fields}

We now consider the configuration in which the first detector is located in the right Rindler wedge and the second in the left wedge, with the interaction involving a right-traveling (RTW) and a left-traveling (LTW) component of the field.

The contribution \(RL_{11}\), involving both interactions on the same detector in the right wedge, remains unchanged from the same-wedge case and is given by (\ref{RL11.same}). However, the cross-term \(RL_{12}\), in which the second detector lies in the left wedge, takes a distinct form due to the change in light-cone coordinate domains. Then \(RL_{12}\) becomes:
\begin{align}
RL_{12} 
=& \left( -\frac{ig}{\hbar} \right)^2 
\int_{-\infty}^{0} du \, \frac{\partial}{\partial u} 
\Phi_{\text{RTW}}(u)
\left( \sigma_1(-a_1 u)^{\frac{i \omega_1}{a_1}} + \sigma_1^\dagger (-a_1 u)^{-\frac{i \omega_1}{a_1}} \right) \nonumber \\
& \phantom{\left( -\frac{ig}{\hbar} \right)^2}
\times \int_{-\infty}^{g(u)} dv \, \frac{\partial}{\partial v} 
\Phi_{\text{LTW}}(v)
\left( \sigma_2(-a_2 v)^{\frac{i \omega_2}{a_2}} + \sigma_2^\dagger (-a_2 v)^{-\frac{i \omega_2}{a_2}} \right). \label{RL12.op}
\end{align}
Here, \(u < 0\) corresponds to the light-cone coordinate of the first detector in the right wedge, and \(v < 0\) denotes the light-cone coordinate of the second detector in the left wedge. To preserve causal time ordering, the upper limit of the second integral must be taken as the negative root \(g(u)\) from equation (\ref{2 in terms of 1}), consistent with the location of the second detector.

Next, $RL_{21}$, with again the first detector lies on the right and the second detector lies on the left wedges, reads
\begin{align}
RL_{21} 
=& \left( -\frac{ig}{\hbar} \right)^2 
\int_{0}^{\infty} du \, \frac{\partial}{\partial u} 
\Phi_{\text{RTW}}(u)
\left( \sigma_2(a_2 u)^{-\frac{i \omega_2}{a_2}} 
+ \sigma_2^\dagger (a_2 u)^{\frac{i \omega_2}{a_2}} \right) \nonumber \\
&\phantom{\left( -\frac{ig}{\hbar} \right)^2}
\times \int_{0}^{h(u)} dv \, \frac{\partial}{\partial v} 
\Phi_{\text{LTW}}(v)
\left( \sigma_1(a_1 v)^{-\frac{i \omega_1}{a_1}} 
+ \sigma_1^\dagger (a_1 v)^{\frac{i \omega_1}{a_1}} \right). \label{RL21.op}
\end{align}
where the positive root \(h(u)\) from (\ref{1 in terms of 2.vu}) is selected to ensure that \(v > 0\). 

Finally the last term $RL_{22}$ becomes
\begin{align}
RL_{22} 
=& \left( -\frac{ig}{\hbar} \right)^2 
\int_{0}^{\infty} du \, \frac{\partial}{\partial u} 
\Phi_{\text{RTW}}(u)
\left( \sigma_2(a_2 u)^{-\frac{i \omega_2}{a_2}} 
+ \sigma_2^\dagger (a_2 u)^{\frac{i \omega_2}{a_2}} \right) \nonumber \\
&\phantom{\left( -\frac{ig}{\hbar} \right)^2}
\times \int_{-\infty}^{-\frac{1}{a_2^2 u}} dv \, \frac{\partial}{\partial v} 
\Phi_{\text{LTW}}(v)
\left( \sigma_2(-a_2 v)^{\frac{i \omega_2}{a_2}} 
+ \sigma_2^\dagger (-a_2 v)^{-\frac{i \omega_2}{a_2}} \right). \label{RL22.op}
\end{align}

\section{Higher-order contributions} \label{higher}

\subsection{Third order}

We now analyze the third-order contribution in the Dyson expansion for the case of two observers (\(m=2\)) interacting locally with a massless scalar field in \(1+1\) dimensions. At each interaction vertex, the field couples either to a right-traveling wave (RTW) or a left-traveling wave (LTW), defined in terms of light-cone coordinates \(u = t - x\) and \(v = t + x\). As a result, the third-order term decomposes into eight distinct contributions according to the mode structure at each interaction time:
\begin{align}
S_{2,3} 
=&\, S_{2,3}^{RRR} 
+ S_{2,3}^{RRL} 
+ S_{2,3}^{RLR} 
+ S_{2,3}^{RLL} \nonumber \\
&+ S_{2,3}^{LRR} 
+ S_{2,3}^{LRL} 
+ S_{2,3}^{LLR} 
+ S_{2,3}^{LLL}.
\end{align}
Each superscript denotes the type of field component (R or L) at the ordered interaction times \(t_1 > t_2 > t_3\). In the following, we compute representative contributions with the detector sequence fixed as: detector 1 at \(t_1\), and detector 2 at \(t_2\) and \(t_3\). Since there are $2^3 = 8$ possible field-mode combinations and $2^3 = 8$ possible detector assignments, the full third-order expansion contains $8 \times 8 = 64$ distinct terms and we fully address all of these terms in the appendix \ref{64}.

\subsection{Finite observers at arbitrary Dyson orders} \label{Dyson.n}

We now consider the most general case involving \(m\) observers, each interacting with a quantum field. Our goal is to evaluate the contribution at the \(n^{\text{th}}\) order in the Dyson expansion. This contribution is denoted by
\begin{align}
\boxed{\quad 
S_{m,n} \equiv \left( \frac{-i}{\hbar} \right)^n 
\int_{-\infty}^{+\infty} dt_1 \sum_{i_1=1}^{m} H_{i_1}(t_1) 
\int_{-\infty}^{t_1} dt_2 \sum_{i_2=1}^{m} H_{i_2}(t_2) 
\dots  
\int_{-\infty}^{t_{n-1}} dt_n \sum_{i_n=1}^{m} H_{i_n}(t_n) 
\quad}\,.
\end{align}

Here, the integration variables \(t_1, \dots, t_n\) correspond to the global Killing time and are temporally ordered as \(t_1 > t_2 > \dots > t_n\), as required by the causal structure of the Dyson series. At each time step \(t_k\), the sum over \(i_k\) runs over all \(m\) observers, incorporating all possible interaction events.

Each observer's Hamiltonian \(H_{i_k}(t_k)\) may be written in terms of their proper time \(\tau_{i_k}(t_k)\), which is assumed to be a monotonic function of the global time \(t_k\). This allows all contributions to be consistently expressed in terms of a single global time coordinate. Hence, at each order \(k\) of the Dyson series, the field-observer interactions are governed entirely by the global time slicing defined by the Killing field.

\section{Conclusion and Outlook} \label{conc}

In this work, we examined the problem of coupling multiple localized detectors to a quantum field in curved spacetime, focusing on the nontrivial role of time-ordering in the Dyson expansion. Using the Unruh--DeWitt detector model in $(1+1)$-dimensional spacetime, we treated the detectors as pointlike classical observers interacting locally with a real scalar Klein-Gordon field.

By working in light-cone coordinates $(u, v)$ and assuming the existence of a timelike Killing vector, we parametrized the interactions along detector trajectories and developed a method to consistently enforce time-ordering in the Dyson series. To handle detectors with differing accelerations, we introduced the functions $f(z), g(z), h(z), k(z)$, which encode the causal structure of time evolution between detectors via coordinate inversion relations.

We classified all second-order Dyson terms and extended the analysis to the full third-order expansion, which contains 64 distinct contributions. Each term was labeled by a mode sequence—such as $RRL_{ijk}$ or $LRL_{ijk}$—and constructed explicitly with integration limits that respect causal ordering. This systematic approach enables a clear and algorithmic treatment of arbitrarily many detectors at higher perturbative orders.

Our results pave the way for several future directions. One immediate extension is to compute the explicit response functions or transition probabilities associated with selected third-order contributions, particularly in entanglement harvesting or information-theoretic settings. Another avenue involves applying the formalism to time-dependent or cosmological spacetimes, where no global Killing time exists, and local approximations to time-ordering become essential. Further generalizations may also include field-theoretic backreaction on the detectors or multi-level detector systems.

\section*{Acknowledgments}

I am grateful to Steve Fulling, Marlan Scully, and Bill Unruh for many illuminating discussions. This work was supported by the
Robert A. Welch Foundation (Grant No. A-1261) and the National Science Foundation (Grant No. PHY-2013771).
\appendix 

\section{64 Terms} \label{64}
Here in this appendix, we present all 64 terms arise in the third order Dyson's series, for a quantum field coupling with two detectors. Here we assume both of detectors are in the right Rindler wedge.

\subsection*{Notation}
In the third-order Dyson expansion, each term can be labeled using the notation 
\[
XYZ_{ijk},
\]
where:
\begin{itemize}
    \item $X, Y, Z \in \{R, L\}$ denote the type of field mode (right-traveling wave $R$ or left-traveling wave $L$) involved in the interaction at time steps $t_1 > t_2 > t_3$, respectively.
    \item The subscripts $i, j, k \in \{1,2\}$ label the observers (e.g., Unruh--DeWitt detectors) that couple to the field at time $t_1$, $t_2$, and $t_3$, respectively.    
\end{itemize}
 For example, the term $RRL_{121}$ corresponds to a contribution where, a right-traveling wave interacts with detector 1 at time $t_1$, another right-traveling wave interacts with detector 2 at time $t_2$, and finally left-traveling wave interacts with detector 1 at time $t_3$.

This labeling convention enables the systematic classification of all third-order contributions.

\subsection*{Detector operator \texorpdfstring{$\Psi_i$}{} in right and left Rindler wedges}

The detector operator $\Psi_i$ corresponds to the interaction of the $i^{\text{th}}$ Unruh--DeWitt detector with the quantum field along its worldline. It is constructed from the detector's internal monopole moment and its proper time evolution.

\subsubsection*{(i) Detector in the right Rindler wedge:}

In the right wedge, the detector's trajectory is parametrized by proper time $\tau$, related to the light-cone coordinates via
\[
-au = e^{-a\tau}, \qquad av = e^{a\tau},
\]
where $u = t - x$ and $v = t + x$. The detector operator takes the form
\[
\Psi_i(u) = \sigma_i \, (-a_i u)^{i \omega_i / a_i} + \sigma_i^\dagger \, (-a_i u)^{-i \omega_i / a_i},
\]
for interactions with right-traveling modes, and
\[
\Psi_i(v) = \sigma_i \, (a_i v)^{-i \omega_i / a_i} + \sigma_i^\dagger \, (a_i v)^{i \omega_i / a_i},
\]
for interactions with left-traveling modes.
\subsubsection*{(ii) Detector in the left Rindler wedge:}

In the left wedge, the detector's proper time relates to light-cone coordinates via
\[
a u = e^{a\tau}, \qquad -a v = e^{-a\tau}.
\]
Accordingly, the detector operator becomes
\[
\Psi_i(u) = \sigma_i \, (a_i u)^{-i \omega_i / a_i} + \sigma_i^\dagger \, (a_i u)^{i \omega_i / a_i},
\]
for right-traveling wave interactions, and
\[
\Psi_i(v) = \sigma_i \, (-a_i v)^{i \omega_i / a_i} + \sigma_i^\dagger \, (-a_i v)^{-i \omega_i / a_i},
\]
for left-traveling wave interactions.

\vspace{0.3cm}

Here, $\sigma_i = \ket{g_i}\bra{e_i}$ is the lowering operator for the $i^{\text{th}}$ detector, and $\omega_i$ denotes the energy gap between its ground and excited states.

\subsection{Third-Order Dyson Terms: \texorpdfstring{$RRR_{ijk}$}{}}

\begin{align}
RRR_{111} 
=& \left( -\frac{ig}{\hbar} \right)^3 
\int_{-\infty}^0 du_1\, \partial_{u_1} \Phi_{\text{RTW}}(u_1) \Psi_1(u_1)
\int_{-\infty}^{u_1} du_2\, \partial_{u_2} \Phi_{\text{RTW}}(u_2) \Psi_1(u_2) \nonumber \\
& \phantom{ \left( -\frac{ig}{\hbar} \right)^3  \quad} \times \int_{-\infty}^{u_2} du_3\, \partial_{u_3} \Phi_{\text{RTW}}(u_3) \Psi_1(u_3) \label{RRR111}
\end{align}

\begin{align}
RRR_{112} 
=& \left( -\frac{ig}{\hbar} \right)^3 
\int_{-\infty}^0 du_1\, \partial_{u_1} \Phi_{\text{RTW}}(u_1) \Psi_1(u_1)
\int_{-\infty}^{u_1} du_2\, \partial_{u_2} \Phi_{\text{RTW}}(u_2) \Psi_1(u_2) \nonumber \\
& \phantom{ \left( -\frac{ig}{\hbar} \right)^3  \quad} 
\times \int_{-\infty}^{g(u_2)} du_3\, \partial_{u_3} \Phi_{\text{RTW}}(u_3) \Psi_2(u_3) \label{RRR112}
\end{align}

\begin{align}
RRR_{121} 
=& \left( -\frac{ig}{\hbar} \right)^3 
\int_{-\infty}^0 du_1\, \partial_{u_1} \Phi_{\text{RTW}}(u_1) \Psi_1(u_1)
\int_{-\infty}^{g(u_1)} du_2\, \partial_{u_2} \Phi_{\text{RTW}}(u_2) \Psi_2(u_2) \nonumber \\
& \phantom{ \left( -\frac{ig}{\hbar} \right)^3  \quad} \times \int_{-\infty}^{k(u_2)} du_3\, \partial_{u_3} \Phi_{\text{RTW}}(u_3) \Psi_1(u_3) \label{RRR121}
\end{align}

\begin{align}
RRR_{122} 
=& \left( -\frac{ig}{\hbar} \right)^3 
\int_{-\infty}^0 du_1\, \partial_{u_1} \Phi_{\text{RTW}}(u_1) \Psi_1(u_1)
\int_{-\infty}^{g(u_1)} du_2\, \partial_{u_2} \Phi_{\text{RTW}}(u_2) \Psi_2(u_2) \nonumber \\
& \phantom{ \left( -\frac{ig}{\hbar} \right)^3  \quad} \times \int_{-\infty}^{g(u_2)} du_3\, \partial_{u_3} \Phi_{\text{RTW}}(u_3) \Psi_2(u_3) \label{RRR122}
\end{align}

\begin{align}
RRR_{211} 
=& \left( -\frac{ig}{\hbar} \right)^3 
\int_{-\infty}^0 du_1\, \partial_{u_1} \Phi_{\text{RTW}}(u_1) \Psi_2(u_1)
\int_{-\infty}^{k(u_1)} du_2\, \partial_{u_2} \Phi_{\text{RTW}}(u_2) \Psi_1(u_2) \nonumber \\
& \phantom{ \left( -\frac{ig}{\hbar} \right)^3  \quad} \times \int_{-\infty}^{u_2} du_3\, \partial_{u_3} \Phi_{\text{RTW}}(u_3) \Psi_1(u_3) \label{RRR211}
\end{align}

\begin{align}
RRR_{212} 
=& \left( -\frac{ig}{\hbar} \right)^3 
\int_{-\infty}^0 du_1\, \partial_{u_1} \Phi_{\text{RTW}}(u_1) \Psi_2(u_1)
\int_{-\infty}^{k(u_1)} du_2\, \partial_{u_2} \Phi_{\text{RTW}}(u_2) \Psi_1(u_2) \nonumber \\
& \phantom{ \left( -\frac{ig}{\hbar} \right)^3  \quad} \times \int_{-\infty}^{g(u_2)} du_3\, \partial_{u_3} \Phi_{\text{RTW}}(u_3) \Psi_2(u_3) \label{RRR212}
\end{align}

\begin{align}
RRR_{221} 
=& \left( -\frac{ig}{\hbar} \right)^3 
\int_{-\infty}^0 du_1\, \partial_{u_1} \Phi_{\text{RTW}}(u_1) \Psi_2(u_1)
\int_{-\infty}^{u_1} du_2\, \partial_{u_2} \Phi_{\text{RTW}}(u_2) \Psi_2(u_2) \nonumber \\
& \phantom{ \left( -\frac{ig}{\hbar} \right)^3  \quad} \times \int_{-\infty}^{k(u_2)} du_3\, \partial_{u_3} \Phi_{\text{RTW}}(u_3) \Psi_1(u_3) \label{RRR221}
\end{align}

\begin{align}
RRR_{222} 
=& \left( -\frac{ig}{\hbar} \right)^3 
\int_{-\infty}^0 du_1\, \partial_{u_1} \Phi_{\text{RTW}}(u_1) \Psi_2(u_1)
\int_{-\infty}^{u_1} du_2\, \partial_{u_2} \Phi_{\text{RTW}}(u_2) \Psi_2(u_2) \nonumber \\
& \phantom{ \left( -\frac{ig}{\hbar} \right)^3  \quad} \times \int_{-\infty}^{u_2} du_3\, \partial_{u_3} \Phi_{\text{RTW}}(u_3) \Psi_2(u_3) \label{RRR222}
\end{align}

\subsection{Third-Order Dyson Terms: \texorpdfstring{$RRL_{ijk}$}{}}

\begin{align}
RRL_{111} 
=& \left( -\frac{ig}{\hbar} \right)^3 
\int_{-\infty}^0 du_1\, \partial_{u_1} \Phi_{\text{RTW}}(u_1) \Psi_1(u_1)
\int_{-\infty}^{u_1} du_2\, \partial_{u_2} \Phi_{\text{RTW}}(u_2) \Psi_1(u_2) \nonumber \\
& \phantom{ \left( -\frac{ig}{\hbar} \right)^3  \quad} \times \int_{0}^{-\frac{1}{a_1^2 u_2}} dv_3\, \partial_{v_3} \Phi_{\text{LTW}}(v_3) \Psi_1(v_3) \label{RRL111}
\end{align}

\begin{align}
RRL_{112} 
=& \left( -\frac{ig}{\hbar} \right)^3 
\int_{-\infty}^0 du_1\, \partial_{u_1} \Phi_{\text{RTW}}(u_1) \Psi_1(u_1)
\int_{-\infty}^{u_1} du_2\, \partial_{u_2} \Phi_{\text{RTW}}(u_2) \Psi_1(u_2) \nonumber \\
& \phantom{ \left( -\frac{ig}{\hbar} \right)^3  \quad} \times \int_{0}^{f(u_2)} dv_3\, \partial_{v_3} \Phi_{\text{LTW}}(v_3) \Psi_2(v_3) \label{RRL112}
\end{align}

\begin{align}
RRL_{121} 
=& \left( -\frac{ig}{\hbar} \right)^3 
\int_{-\infty}^0 du_1\, \partial_{u_1} \Phi_{\text{RTW}}(u_1) \Psi_1(u_1)
\int_{-\infty}^{g(u_1)} du_2\, \partial_{u_2} \Phi_{\text{RTW}}(u_2) \Psi_2(u_2) \nonumber \\
& \phantom{ \left( -\frac{ig}{\hbar} \right)^3  \quad} \times \int_{0}^{h(u_2)} dv_3\, \partial_{v_3} \Phi_{\text{LTW}}(v_3) \Psi_1(v_3) \label{RRL121}
\end{align}

\begin{align}
RRL_{122} 
=& \left( -\frac{ig}{\hbar} \right)^3 
\int_{-\infty}^0 du_1\, \partial_{u_1} \Phi_{\text{RTW}}(u_1) \Psi_1(u_1)
\int_{-\infty}^{g(u_1)} du_2\, \partial_{u_2} \Phi_{\text{RTW}}(u_2) \Psi_2(u_2) \nonumber \\
& \phantom{ \left( -\frac{ig}{\hbar} \right)^3  \quad} \times \int_{0}^{-\frac{1}{a_2^2 u_2}} dv_3\, \partial_{v_3} \Phi_{\text{LTW}}(v_3) \Psi_2(v_3) \label{RRL122}
\end{align}

\begin{align}
RRL_{211} 
=& \left( -\frac{ig}{\hbar} \right)^3 
\int_{-\infty}^0 du_1\, \partial_{u_1} \Phi_{\text{RTW}}(u_1) \Psi_2(u_1)
\int_{-\infty}^{k(u_1)} du_2\, \partial_{u_2} \Phi_{\text{RTW}}(u_2) \Psi_1(u_2) \nonumber \\
& \phantom{ \left( -\frac{ig}{\hbar} \right)^3  \quad} \times \int_{0}^{-\frac{1}{a_1^2 u_2}} dv_3\, \partial_{v_3} \Phi_{\text{LTW}}(v_3) \Psi_1(v_3) \label{RRL211}
\end{align}

\begin{align}
RRL_{212} 
=& \left( -\frac{ig}{\hbar} \right)^3 
\int_{-\infty}^0 du_1\, \partial_{u_1} \Phi_{\text{RTW}}(u_1) \Psi_2(u_1)
\int_{-\infty}^{k(u_1)} du_2\, \partial_{u_2} \Phi_{\text{RTW}}(u_2) \Psi_1(u_2) \nonumber \\
& \phantom{ \left( -\frac{ig}{\hbar} \right)^3  \quad} \times \int_{0}^{f(u_2)} dv_3\, \partial_{v_3} \Phi_{\text{LTW}}(v_3) \Psi_2(v_3) \label{RRL212}
\end{align}

\begin{align}
RRL_{221} 
=& \left( -\frac{ig}{\hbar} \right)^3 
\int_{-\infty}^0 du_1\, \partial_{u_1} \Phi_{\text{RTW}}(u_1) \Psi_2(u_1)
\int_{-\infty}^{u_1} du_2\, \partial_{u_2} \Phi_{\text{RTW}}(u_2) \Psi_2(u_2) \nonumber \\
& \phantom{ \left( -\frac{ig}{\hbar} \right)^3  \quad} \times \int_{0}^{h(u_2)} dv_3\, \partial_{v_3} \Phi_{\text{LTW}}(v_3) \Psi_1(v_3) \label{RRL221}
\end{align}

\begin{align}
RRL_{222} 
=& \left( -\frac{ig}{\hbar} \right)^3 
\int_{-\infty}^0 du_1\, \partial_{u_1} \Phi_{\text{RTW}}(u_1) \Psi_2(u_1)
\int_{-\infty}^{u_1} du_2\, \partial_{u_2} \Phi_{\text{RTW}}(u_2) \Psi_2(u_2) \nonumber \\
& \phantom{ \left( -\frac{ig}{\hbar} \right)^3  \quad} \times \int_{0}^{-\frac{1}{a_2^2 u_2}} dv_3\, \partial_{v_3} \Phi_{\text{LTW}}(v_3) \Psi_2(v_3) \label{RRL222}
\end{align}

\subsection{Third-Order Dyson Terms: \texorpdfstring{$RLR_{ijk}$}{}}

\begin{align}
RLR_{111} 
=& \left( -\frac{ig}{\hbar} \right)^3 
\int_{-\infty}^0 du_1\, \partial_{u_1} \Phi_{\text{RTW}}(u_1) \Psi_1(u_1)
\int_{0}^{-\frac{1}{a_1^2 u_1}} dv_2\, \partial_{v_2} \Phi_{\text{LTW}}(v_2) \Psi_1(v_2) \nonumber \\
& \phantom{ \left( -\frac{ig}{\hbar} \right)^3  \quad} \times \int_{-\infty}^{-\frac{1}{a_1^2 v_2}} du_3\, \partial_{u_3} \Phi_{\text{RTW}}(u_3) \Psi_1(u_3) \label{RLR111}
\end{align}

\begin{align}
RLR_{112} 
=& \left( -\frac{ig}{\hbar} \right)^3 
\int_{-\infty}^0 du_1\, \partial_{u_1} \Phi_{\text{RTW}}(u_1) \Psi_1(u_1)
\int_{0}^{-\frac{1}{a_1^2 u_1}} dv_2\, \partial_{v_2} \Phi_{\text{LTW}}(v_2) \Psi_1(v_2) \nonumber \\
& \phantom{ \left( -\frac{ig}{\hbar} \right)^3  \quad} \times \int_{-\infty}^{g(v_2)} du_3\, \partial_{u_3} \Phi_{\text{RTW}}(u_3) \Psi_2(u_3) \label{RLR112}
\end{align}

\begin{align}
RLR_{121} 
=& \left( -\frac{ig}{\hbar} \right)^3 
\int_{-\infty}^0 du_1\, \partial_{u_1} \Phi_{\text{RTW}}(u_1) \Psi_1(u_1)
\int_{0}^{f(u_1)} dv_2\, \partial_{v_2} \Phi_{\text{LTW}}(v_2) \Psi_2(v_2) \nonumber \\
& \phantom{ \left( -\frac{ig}{\hbar} \right)^3  \quad} \times \int_{-\infty}^{k(v_2)} du_3\, \partial_{u_3} \Phi_{\text{RTW}}(u_3) \Psi_1(u_3) \label{RLR121}
\end{align}

\begin{align}
RLR_{122} 
=& \left( -\frac{ig}{\hbar} \right)^3 
\int_{-\infty}^0 du_1\, \partial_{u_1} \Phi_{\text{RTW}}(u_1) \Psi_1(u_1)
\int_{0}^{f(u_1)} dv_2\, \partial_{v_2} \Phi_{\text{LTW}}(v_2) \Psi_2(v_2) \nonumber \\
& \phantom{ \left( -\frac{ig}{\hbar} \right)^3  \quad} \times \int_{-\infty}^{-\frac{1}{a_2^2 v_2}} du_3\, \partial_{u_3} \Phi_{\text{RTW}}(u_3) \Psi_2(u_3) \label{RLR122}
\end{align}

\begin{align}
RLR_{211} 
=& \left( -\frac{ig}{\hbar} \right)^3 
\int_{-\infty}^0 du_1\, \partial_{u_1} \Phi_{\text{RTW}}(u_1) \Psi_2(u_1)
\int_{0}^{h(u_1)} dv_2\, \partial_{v_2} \Phi_{\text{LTW}}(v_2) \Psi_1(v_2) \nonumber \\
& \phantom{ \left( -\frac{ig}{\hbar} \right)^3  \quad} \times \int_{-\infty}^{-\frac{1}{a_1^2 v_2}} du_3\, \partial_{u_3} \Phi_{\text{RTW}}(u_3) \Psi_1(u_3) \label{RLR211}
\end{align}

\begin{align}
RLR_{212} 
=& \left( -\frac{ig}{\hbar} \right)^3 
\int_{-\infty}^0 du_1\, \partial_{u_1} \Phi_{\text{RTW}}(u_1) \Psi_2(u_1)
\int_{0}^{h(u_1)} dv_2\, \partial_{v_2} \Phi_{\text{LTW}}(v_2) \Psi_1(v_2) \nonumber \\
& \phantom{ \left( -\frac{ig}{\hbar} \right)^3  \quad} \times \int_{-\infty}^{g(v_2)} du_3\, \partial_{u_3} \Phi_{\text{RTW}}(u_3) \Psi_2(u_3) \label{RLR212}
\end{align}

\begin{align}
RLR_{221} 
=& \left( -\frac{ig}{\hbar} \right)^3 
\int_{-\infty}^0 du_1\, \partial_{u_1} \Phi_{\text{RTW}}(u_1) \Psi_2(u_1)
\int_{0}^{-\frac{1}{a_2^2 u_1}} dv_2\, \partial_{v_2} \Phi_{\text{LTW}}(v_2) \Psi_2(v_2) \nonumber \\
& \phantom{ \left( -\frac{ig}{\hbar} \right)^3  \quad} \times \int_{-\infty}^{k(v_2)} du_3\, \partial_{u_3} \Phi_{\text{RTW}}(u_3) \Psi_1(u_3) \label{RLR221}
\end{align}

\begin{align}
RLR_{222} 
=& \left( -\frac{ig}{\hbar} \right)^3 
\int_{-\infty}^0 du_1\, \partial_{u_1} \Phi_{\text{RTW}}(u_1) \Psi_2(u_1)
\int_{0}^{-\frac{1}{a_2^2 u_1}} dv_2\, \partial_{v_2} \Phi_{\text{LTW}}(v_2) \Psi_2(v_2) \nonumber \\
& \phantom{ \left( -\frac{ig}{\hbar} \right)^3  \quad} \times \int_{-\infty}^{-\frac{1}{a_2^2 v_2}} du_3\, \partial_{u_3} \Phi_{\text{RTW}}(u_3) \Psi_2(u_3) \label{RLR222}
\end{align}

\subsection{Third-Order Dyson Terms: \texorpdfstring{$RLL_{ijk}$}{}}

\begin{align}
RLL_{111} 
=& \left( -\frac{ig}{\hbar} \right)^3 
\int_{-\infty}^0 du_1\, \partial_{u_1} \Phi_{\text{RTW}}(u_1) \Psi_1(u_1)
\int_0^{- \frac{1}{a_1^2 u_1}} dv_2\, \partial_{v_2} \Phi_{\text{LTW}}(v_2) \Psi_1(v_2) \nonumber \\
& \phantom{ \left( -\frac{ig}{\hbar} \right)^3  \quad} \times \int_0^{- \frac{1}{a_1^2 v_2}} dv_3\, \partial_{v_3} \Phi_{\text{LTW}}(v_3) \Psi_1(v_3) \label{RLL111}
\end{align}

\begin{align}
RLL_{112} 
=& \left( -\frac{ig}{\hbar} \right)^3 
\int_{-\infty}^0 du_1\, \partial_{u_1} \Phi_{\text{RTW}}(u_1) \Psi_1(u_1)
\int_0^{- \frac{1}{a_1^2 u_1}} dv_2\, \partial_{v_2} \Phi_{\text{LTW}}(v_2) \Psi_1(v_2) \nonumber \\
& \phantom{ \left( -\frac{ig}{\hbar} \right)^3  \quad} \times \int_0^{f(v_2)} dv_3\, \partial_{v_3} \Phi_{\text{LTW}}(v_3) \Psi_2(v_3) \label{RLL112}
\end{align}

\begin{align}
RLL_{121} 
=& \left( -\frac{ig}{\hbar} \right)^3 
\int_{-\infty}^0 du_1\, \partial_{u_1} \Phi_{\text{RTW}}(u_1) \Psi_1(u_1)
\int_0^{f(u_1)} dv_2\, \partial_{v_2} \Phi_{\text{LTW}}(v_2) \Psi_2(v_2) \nonumber \\
& \phantom{ \left( -\frac{ig}{\hbar} \right)^3  \quad} \times \int_0^{h(v_2)} dv_3\, \partial_{v_3} \Phi_{\text{LTW}}(v_3) \Psi_1(v_3) \label{RLL121}
\end{align}

\begin{align}
RLL_{122} 
=& \left( -\frac{ig}{\hbar} \right)^3 
\int_{-\infty}^0 du_1\, \partial_{u_1} \Phi_{\text{RTW}}(u_1) \Psi_1(u_1)
\int_0^{f(u_1)} dv_2\, \partial_{v_2} \Phi_{\text{LTW}}(v_2) \Psi_2(v_2) \nonumber \\
& \phantom{ \left( -\frac{ig}{\hbar} \right)^3  \quad} \times \int_0^{-\frac{1}{a_2^2 v_2}} dv_3\, \partial_{v_3} \Phi_{\text{LTW}}(v_3) \Psi_2(v_3) \label{RLL122}
\end{align}

\begin{align}
RLL_{211} 
=& \left( -\frac{ig}{\hbar} \right)^3 
\int_{-\infty}^0 du_1\, \partial_{u_1} \Phi_{\text{RTW}}(u_1) \Psi_2(u_1)
\int_0^{h(u_1)} dv_2\, \partial_{v_2} \Phi_{\text{LTW}}(v_2) \Psi_1(v_2) \nonumber \\
& \phantom{ \left( -\frac{ig}{\hbar} \right)^3  \quad} \times \int_0^{- \frac{1}{a_1^2 v_2}} dv_3\, \partial_{v_3} \Phi_{\text{LTW}}(v_3) \Psi_1(v_3) \label{RLL211}
\end{align}

\begin{align}
RLL_{212} 
=& \left( -\frac{ig}{\hbar} \right)^3 
\int_{-\infty}^0 du_1\, \partial_{u_1} \Phi_{\text{RTW}}(u_1) \Psi_2(u_1)
\int_0^{h(u_1)} dv_2\, \partial_{v_2} \Phi_{\text{LTW}}(v_2) \Psi_1(v_2) \nonumber \\
& \phantom{ \left( -\frac{ig}{\hbar} \right)^3  \quad} \times \int_0^{f(v_2)} dv_3\, \partial_{v_3} \Phi_{\text{LTW}}(v_3) \Psi_2(v_3) \label{RLL212}
\end{align}

\begin{align}
RLL_{221} 
=& \left( -\frac{ig}{\hbar} \right)^3 
\int_{-\infty}^0 du_1\, \partial_{u_1} \Phi_{\text{RTW}}(u_1) \Psi_2(u_1)
\int_0^{- \frac{1}{a_2^2 u_1}} dv_2\, \partial_{v_2} \Phi_{\text{LTW}}(v_2) \Psi_2(v_2) \nonumber \\
& \phantom{ \left( -\frac{ig}{\hbar} \right)^3  \quad} \times \int_0^{h(v_2)} dv_3\, \partial_{v_3} \Phi_{\text{LTW}}(v_3) \Psi_1(v_3) \label{RLL221}
\end{align}

\begin{align}
RLL_{222} 
=& \left( -\frac{ig}{\hbar} \right)^3 
\int_{-\infty}^0 du_1\, \partial_{u_1} \Phi_{\text{RTW}}(u_1) \Psi_2(u_1)
\int_0^{- \frac{1}{a_2^2 u_1}} dv_2\, \partial_{v_2} \Phi_{\text{LTW}}(v_2) \Psi_2(v_2) \nonumber \\
& \phantom{ \left( -\frac{ig}{\hbar} \right)^3  \quad} \times \int_0^{- \frac{1}{a_2^2 v_2}} dv_3\, \partial_{v_3} \Phi_{\text{LTW}}(v_3) \Psi_2(v_3) \label{RLL222}
\end{align}

\subsection{Third-Order Dyson Terms: \texorpdfstring{$LRR_{ijk}$}{}}

\begin{align}
LRR_{111} 
=& \left( -\frac{ig}{\hbar} \right)^3 
\int_{0}^{\infty} dv_1\, \partial_{v_1} \Phi_{\text{LTW}}(v_1) \Psi_1(v_1)
\int_{-\infty}^{-\frac{1}{a_1^2 v_1}} du_2\, \partial_{u_2} \Phi_{\text{RTW}}(u_2) \Psi_1(u_2) \nonumber \\
& \phantom{ \left( -\frac{ig}{\hbar} \right)^3  \quad} \times \int_{-\infty}^{u_2} du_3\, \partial_{u_3} \Phi_{\text{RTW}}(u_3) \Psi_1(u_3) \label{LRR111}
\end{align}

\begin{align}
LRR_{112} 
=& \left( -\frac{ig}{\hbar} \right)^3 
\int_{0}^{\infty} dv_1\, \partial_{v_1} \Phi_{\text{LTW}}(v_1) \Psi_1(v_1)
\int_{-\infty}^{-\frac{1}{a_1^2 v_1}} du_2\, \partial_{u_2} \Phi_{\text{RTW}}(u_2) \Psi_1(u_2) \nonumber \\
& \phantom{ \left( -\frac{ig}{\hbar} \right)^3  \quad} \times \int_{-\infty}^{g(u_2)} du_3\, \partial_{u_3} \Phi_{\text{RTW}}(u_3) \Psi_2(u_3) \label{LRR112}
\end{align}

\begin{align}
LRR_{121} 
=& \left( -\frac{ig}{\hbar} \right)^3 
\int_{0}^{\infty} dv_1\, \partial_{v_1} \Phi_{\text{LTW}}(v_1) \Psi_1(v_1)
\int_{-\infty}^{g(v_1)} du_2\, \partial_{u_2} \Phi_{\text{RTW}}(u_2) \Psi_2(u_2) \nonumber \\
& \phantom{ \left( -\frac{ig}{\hbar} \right)^3  \quad} \times \int_{-\infty}^{k(u_2)} du_3\, \partial_{u_3} \Phi_{\text{RTW}}(u_3) \Psi_1(u_3) \label{LRR121}
\end{align}

\begin{align}
LRR_{122} 
=& \left( -\frac{ig}{\hbar} \right)^3 
\int_{0}^{\infty} dv_1\, \partial_{v_1} \Phi_{\text{LTW}}(v_1) \Psi_1(v_1)
\int_{-\infty}^{g(v_1)} du_2\, \partial_{u_2} \Phi_{\text{RTW}}(u_2) \Psi_2(u_2) \nonumber \\
& \phantom{ \left( -\frac{ig}{\hbar} \right)^3  \quad} \times \int_{-\infty}^{u_2} du_3\, \partial_{u_3} \Phi_{\text{RTW}}(u_3) \Psi_2(u_3) \label{LRR122}
\end{align}

\begin{align}
LRR_{211} 
=& \left( -\frac{ig}{\hbar} \right)^3 
\int_{0}^{\infty} dv_1\, \partial_{v_1} \Phi_{\text{LTW}}(v_1) \Psi_2(v_1)
\int_{-\infty}^{k(v_1)} du_2\, \partial_{u_2} \Phi_{\text{RTW}}(u_2) \Psi_1(u_2) \nonumber \\
& \phantom{ \left( -\frac{ig}{\hbar} \right)^3  \quad} \times \int_{-\infty}^{u_2} du_3\, \partial_{u_3} \Phi_{\text{RTW}}(u_3) \Psi_1(u_3) \label{LRR211}
\end{align}

\begin{align}
LRR_{212} 
=& \left( -\frac{ig}{\hbar} \right)^3 
\int_{0}^{\infty} dv_1\, \partial_{v_1} \Phi_{\text{LTW}}(v_1) \Psi_2(v_1)
\int_{-\infty}^{k(v_1)} du_2\, \partial_{u_2} \Phi_{\text{RTW}}(u_2) \Psi_1(u_2) \nonumber \\
& \phantom{ \left( -\frac{ig}{\hbar} \right)^3  \quad} \times \int_{-\infty}^{g(u_2)} du_3\, \partial_{u_3} \Phi_{\text{RTW}}(u_3) \Psi_2(u_3) \label{LRR212}
\end{align}

\begin{align}
LRR_{221} 
=& \left( -\frac{ig}{\hbar} \right)^3 
\int_{0}^{\infty} dv_1\, \partial_{v_1} \Phi_{\text{LTW}}(v_1) \Psi_2(v_1)
\int_{-\infty}^{-\frac{1}{a_2^2 v_1}} du_2\, \partial_{u_2} \Phi_{\text{RTW}}(u_2) \Psi_2(u_2) \nonumber \\
& \phantom{ \left( -\frac{ig}{\hbar} \right)^3  \quad} \times \int_{-\infty}^{k(u_2)} du_3\, \partial_{u_3} \Phi_{\text{RTW}}(u_3) \Psi_1(u_3) \label{LRR221}
\end{align}

\begin{align}
LRR_{222} 
=& \left( -\frac{ig}{\hbar} \right)^3 
\int_{0}^{\infty} dv_1\, \partial_{v_1} \Phi_{\text{LTW}}(v_1) \Psi_2(v_1)
\int_{-\infty}^{-\frac{1}{a_2^2 v_1}} du_2\, \partial_{u_2} \Phi_{\text{RTW}}(u_2) \Psi_2(u_2) \nonumber \\
& \phantom{ \left( -\frac{ig}{\hbar} \right)^3  \quad} \times \int_{-\infty}^{u_2} du_3\, \partial_{u_3} \Phi_{\text{RTW}}(u_3) \Psi_2(u_3) \label{LRR222}
\end{align}

\subsection{Third-Order Dyson Terms: \texorpdfstring{$LRL_{ijk}$}{}}

\begin{align}
LRL_{111} 
=& \left( -\frac{ig}{\hbar} \right)^3 
\int_0^{\infty} dv_1\, \partial_{v_1} \Phi_{\text{LTW}}(v_1) \Psi_1(v_1)
\int_{-\infty}^{- \frac{1}{a_1^2 v_1}} du_2\, \partial_{u_2} \Phi_{\text{RTW}}(u_2) \Psi_1(u_2) \nonumber \\
& \phantom{ \left( -\frac{ig}{\hbar} \right)^3  \quad} \times \int_0^{- \frac{1}{a_1^2 u_2}} dv_3\, \partial_{v_3} \Phi_{\text{LTW}}(v_3) \Psi_1(v_3) \label{LRL111}
\end{align}

\begin{align}
LRL_{112} 
=& \left( -\frac{ig}{\hbar} \right)^3 
\int_0^{\infty} dv_1\, \partial_{v_1} \Phi_{\text{LTW}}(v_1) \Psi_1(v_1)
\int_{-\infty}^{- \frac{1}{a_1^2 v_1}} du_2\, \partial_{u_2} \Phi_{\text{RTW}}(u_2) \Psi_1(u_2) \nonumber \\
& \phantom{ \left( -\frac{ig}{\hbar} \right)^3  \quad} \times \int_0^{f(u_2)} dv_3\, \partial_{v_3} \Phi_{\text{LTW}}(v_3) \Psi_2(v_3) \label{LRL112}
\end{align}

\begin{align}
LRL_{121} 
=& \left( -\frac{ig}{\hbar} \right)^3 
\int_0^{\infty} dv_1\, \partial_{v_1} \Phi_{\text{LTW}}(v_1) \Psi_1(v_1)
\int_{-\infty}^{g(v_1)} du_2\, \partial_{u_2} \Phi_{\text{RTW}}(u_2) \Psi_2(u_2) \nonumber \\
& \phantom{ \left( -\frac{ig}{\hbar} \right)^3  \quad} \times \int_0^{h(u_2)} dv_3\, \partial_{v_3} \Phi_{\text{LTW}}(v_3) \Psi_1(v_3) \label{LRL121}
\end{align}

\begin{align}
LRL_{122} 
=& \left( -\frac{ig}{\hbar} \right)^3 
\int_0^{\infty} dv_1\, \partial_{v_1} \Phi_{\text{LTW}}(v_1) \Psi_1(v_1)
\int_{-\infty}^{g(v_1)} du_2\, \partial_{u_2} \Phi_{\text{RTW}}(u_2) \Psi_2(u_2) \nonumber \\
& \phantom{ \left( -\frac{ig}{\hbar} \right)^3  \quad} \times \int_0^{-\frac{1}{a_2^2 u_2}} dv_3\, \partial_{v_3} \Phi_{\text{LTW}}(v_3) \Psi_2(v_3) \label{LRL122}
\end{align}

\begin{align}
LRL_{211} 
=& \left( -\frac{ig}{\hbar} \right)^3 
\int_0^{\infty} dv_1\, \partial_{v_1} \Phi_{\text{LTW}}(v_1) \Psi_2(v_1)
\int_{-\infty}^{k(v_1)} du_2\, \partial_{u_2} \Phi_{\text{RTW}}(u_2) \Psi_1(u_2) \nonumber \\
& \phantom{ \left( -\frac{ig}{\hbar} \right)^3  \quad} \times \int_0^{- \frac{1}{a_1^2 u_2}} dv_3\, \partial_{v_3} \Phi_{\text{LTW}}(v_3) \Psi_1(v_3) \label{LRL211}
\end{align}

\begin{align}
LRL_{212} 
=& \left( -\frac{ig}{\hbar} \right)^3 
\int_0^{\infty} dv_1\, \partial_{v_1} \Phi_{\text{LTW}}(v_1) \Psi_2(v_1)
\int_{-\infty}^{k(v_1)} du_2\, \partial_{u_2} \Phi_{\text{RTW}}(u_2) \Psi_1(u_2) \nonumber \\
& \phantom{ \left( -\frac{ig}{\hbar} \right)^3  \quad} \times \int_0^{f(u_2)} dv_3\, \partial_{v_3} \Phi_{\text{LTW}}(v_3) \Psi_2(v_3) \label{LRL212}
\end{align}

\begin{align}
LRL_{221} 
=& \left( -\frac{ig}{\hbar} \right)^3 
\int_0^{\infty} dv_1\, \partial_{v_1} \Phi_{\text{LTW}}(v_1) \Psi_2(v_1)
\int_{-\infty}^{- \frac{1}{a_2^2 v_1}} du_2\, \partial_{u_2} \Phi_{\text{RTW}}(u_2) \Psi_2(u_2) \nonumber \\
& \phantom{ \left( -\frac{ig}{\hbar} \right)^3  \quad} \times \int_0^{h(u_2)} dv_3\, \partial_{v_3} \Phi_{\text{LTW}}(v_3) \Psi_1(v_3) \label{LRL221}
\end{align}

\begin{align}
LRL_{222} 
=& \left( -\frac{ig}{\hbar} \right)^3 
\int_0^{\infty} dv_1\, \partial_{v_1} \Phi_{\text{LTW}}(v_1) \Psi_2(v_1)
\int_{-\infty}^{- \frac{1}{a_2^2 v_1}} du_2\, \partial_{u_2} \Phi_{\text{RTW}}(u_2) \Psi_2(u_2) \nonumber \\
& \phantom{ \left( -\frac{ig}{\hbar} \right)^3  \quad} \times \int_0^{- \frac{1}{a_2^2 u_2}} dv_3\, \partial_{v_3} \Phi_{\text{LTW}}(v_3) \Psi_2(v_3) \label{LRL222}
\end{align}

\subsection{Third-Order Dyson Terms: \texorpdfstring{$LLR_{ijk}$}{}}

\begin{align}
LLR_{111} 
=& \left( -\frac{ig}{\hbar} \right)^3 
\int_0^{\infty} dv_1\, \partial_{v_1} \Phi_{\text{LTW}}(v_1) \Psi_1(v_1)
\int_0^{v_1} dv_2\, \partial_{v_2} \Phi_{\text{LTW}}(v_2) \Psi_1(v_2) \nonumber \\
& \phantom{ \left( -\frac{ig}{\hbar} \right)^3  \quad} \times \int_{-\infty}^{-\frac{1}{a_1^2 v_2}} du_3\, \partial_{u_3} \Phi_{\text{RTW}}(u_3) \Psi_1(u_3) \label{LLR111}
\end{align}

\begin{align}
LLR_{112} 
=& \left( -\frac{ig}{\hbar} \right)^3 
\int_0^{\infty} dv_1\, \partial_{v_1} \Phi_{\text{LTW}}(v_1) \Psi_1(v_1)
\int_0^{v_1} dv_2\, \partial_{v_2} \Phi_{\text{LTW}}(v_2) \Psi_1(v_2) \nonumber \\
& \phantom{ \left( -\frac{ig}{\hbar} \right)^3  \quad} \times \int_{-\infty}^{g(v_2)} du_3\, \partial_{u_3} \Phi_{\text{RTW}}(u_3) \Psi_2(u_3) \label{LLR112}
\end{align}

\begin{align}
LLR_{121} 
=& \left( -\frac{ig}{\hbar} \right)^3 
\int_0^{\infty} dv_1\, \partial_{v_1} \Phi_{\text{LTW}}(v_1) \Psi_1(v_1)
\int_0^{f(v_1)} dv_2\, \partial_{v_2} \Phi_{\text{LTW}}(v_2) \Psi_2(v_2) \nonumber \\
& \phantom{ \left( -\frac{ig}{\hbar} \right)^3  \quad} \times \int_{-\infty}^{k(v_2)} du_3\, \partial_{u_3} \Phi_{\text{RTW}}(u_3) \Psi_1(u_3) \label{LLR121}
\end{align}

\begin{align}
LLR_{122} 
=& \left( -\frac{ig}{\hbar} \right)^3 
\int_0^{\infty} dv_1\, \partial_{v_1} \Phi_{\text{LTW}}(v_1) \Psi_1(v_1)
\int_0^{f(v_1)} dv_2\, \partial_{v_2} \Phi_{\text{LTW}}(v_2) \Psi_2(v_2) \nonumber \\
& \phantom{ \left( -\frac{ig}{\hbar} \right)^3  \quad} \times \int_{-\infty}^{-\frac{1}{a_2^2 v_2}} du_3\, \partial_{u_3} \Phi_{\text{RTW}}(u_3) \Psi_2(u_3) \label{LLR122}
\end{align}

\begin{align}
LLR_{211} 
=& \left( -\frac{ig}{\hbar} \right)^3 
\int_0^{\infty} dv_1\, \partial_{v_1} \Phi_{\text{LTW}}(v_1) \Psi_2(v_1)
\int_0^{h(v_1)} dv_2\, \partial_{v_2} \Phi_{\text{LTW}}(v_2) \Psi_1(v_2) \nonumber \\
& \phantom{ \left( -\frac{ig}{\hbar} \right)^3  \quad} \times \int_{-\infty}^{-\frac{1}{a_1^2 v_2}} du_3\, \partial_{u_3} \Phi_{\text{RTW}}(u_3) \Psi_1(u_3) \label{LLR211}
\end{align}

\begin{align}
LLR_{212} 
=& \left( -\frac{ig}{\hbar} \right)^3 
\int_0^{\infty} dv_1\, \partial_{v_1} \Phi_{\text{LTW}}(v_1) \Psi_2(v_1)
\int_0^{h(v_1)} dv_2\, \partial_{v_2} \Phi_{\text{LTW}}(v_2) \Psi_1(v_2) \nonumber \\
& \phantom{ \left( -\frac{ig}{\hbar} \right)^3  \quad} \times \int_{-\infty}^{g(v_2)} du_3\, \partial_{u_3} \Phi_{\text{RTW}}(u_3) \Psi_2(u_3) \label{LLR212}
\end{align}

\begin{align}
LLR_{221} 
=& \left( -\frac{ig}{\hbar} \right)^3 
\int_0^{\infty} dv_1\, \partial_{v_1} \Phi_{\text{LTW}}(v_1) \Psi_2(v_1)
\int_0^{v_1} dv_2\, \partial_{v_2} \Phi_{\text{LTW}}(v_2) \Psi_2(v_2) \nonumber \\
& \phantom{ \left( -\frac{ig}{\hbar} \right)^3  \quad} \times \int_{-\infty}^{k(v_2)} du_3\, \partial_{u_3} \Phi_{\text{RTW}}(u_3) \Psi_1(u_3) \label{LLR221}
\end{align}

\begin{align}
LLR_{222} 
=& \left( -\frac{ig}{\hbar} \right)^3 
\int_0^{\infty} dv_1\, \partial_{v_1} \Phi_{\text{LTW}}(v_1) \Psi_2(v_1)
\int_0^{v_1} dv_2\, \partial_{v_2} \Phi_{\text{LTW}}(v_2) \Psi_2(v_2) \nonumber \\
& \phantom{ \left( -\frac{ig}{\hbar} \right)^3  \quad} \times \int_{-\infty}^{-\frac{1}{a_2^2 v_2}} du_3\, \partial_{u_3} \Phi_{\text{RTW}}(u_3) \Psi_2(u_3) \label{LLR222}
\end{align}

\subsection{Third-Order Dyson Terms: \texorpdfstring{$LLL_{ijk}$}{}}

\begin{align}
LLL_{111} 
=& \left( -\frac{ig}{\hbar} \right)^3 
\int_{0}^{\infty} dv_1\, \partial_{v_1} \Phi_{\text{LTW}}(v_1) \Psi_1(v_1)
\int_{0}^{v_1} dv_2\, \partial_{v_2} \Phi_{\text{LTW}}(v_2) \Psi_1(v_2) \nonumber \\
& \phantom{ \left( -\frac{ig}{\hbar} \right)^3  \quad} \times \int_{0}^{v_2} dv_3\, \partial_{v_3} \Phi_{\text{LTW}}(v_3) \Psi_1(v_3) \label{LLL111}
\end{align}

\begin{align}
LLL_{112} 
=& \left( -\frac{ig}{\hbar} \right)^3 
\int_{0}^{\infty} dv_1\, \partial_{v_1} \Phi_{\text{LTW}}(v_1) \Psi_1(v_1)
\int_{0}^{v_1} dv_2\, \partial_{v_2} \Phi_{\text{LTW}}(v_2) \Psi_1(v_2) \nonumber \\
& \phantom{ \left( -\frac{ig}{\hbar} \right)^3  \quad} \times \int_{0}^{f(v_2)} dv_3\, \partial_{v_3} \Phi_{\text{LTW}}(v_3) \Psi_2(v_3) \label{LLL112}
\end{align}

\begin{align}
LLL_{121} 
=& \left( -\frac{ig}{\hbar} \right)^3 
\int_{0}^{\infty} dv_1\, \partial_{v_1} \Phi_{\text{LTW}}(v_1) \Psi_1(v_1)
\int_{0}^{f(v_1)} dv_2\, \partial_{v_2} \Phi_{\text{LTW}}(v_2) \Psi_2(v_2) \nonumber \\
& \phantom{ \left( -\frac{ig}{\hbar} \right)^3  \quad} \times \int_{0}^{h(v_2)} dv_3\, \partial_{v_3} \Phi_{\text{LTW}}(v_3) \Psi_1(v_3) \label{LLL121}
\end{align}

\begin{align}
LLL_{122} 
=& \left( -\frac{ig}{\hbar} \right)^3 
\int_{0}^{\infty} dv_1\, \partial_{v_1} \Phi_{\text{LTW}}(v_1) \Psi_1(v_1)
\int_{0}^{f(v_1)} dv_2\, \partial_{v_2} \Phi_{\text{LTW}}(v_2) \Psi_2(v_2) \nonumber \\
& \phantom{ \left( -\frac{ig}{\hbar} \right)^3  \quad} \times \int_{0}^{v_2} dv_3\, \partial_{v_3} \Phi_{\text{LTW}}(v_3) \Psi_2(v_3) \label{LLL122}
\end{align}

\begin{align}
LLL_{211} 
=& \left( -\frac{ig}{\hbar} \right)^3 
\int_{0}^{\infty} dv_1\, \partial_{v_1} \Phi_{\text{LTW}}(v_1) \Psi_2(v_1)
\int_{0}^{h(v_1)} dv_2\, \partial_{v_2} \Phi_{\text{LTW}}(v_2) \Psi_1(v_2) \nonumber \\
& \phantom{ \left( -\frac{ig}{\hbar} \right)^3  \quad} \times \int_{0}^{v_2} dv_3\, \partial_{v_3} \Phi_{\text{LTW}}(v_3) \Psi_1(v_3) \label{LLL211}
\end{align}

\begin{align}
LLL_{212} 
=& \left( -\frac{ig}{\hbar} \right)^3 
\int_{0}^{\infty} dv_1\, \partial_{v_1} \Phi_{\text{LTW}}(v_1) \Psi_2(v_1)
\int_{0}^{h(v_1)} dv_2\, \partial_{v_2} \Phi_{\text{LTW}}(v_2) \Psi_1(v_2) \nonumber \\
& \phantom{ \left( -\frac{ig}{\hbar} \right)^3  \quad} \times \int_{0}^{f(v_2)} dv_3\, \partial_{v_3} \Phi_{\text{LTW}}(v_3) \Psi_2(v_3) \label{LLL212}
\end{align}

\begin{align}
LLL_{221} 
=& \left( -\frac{ig}{\hbar} \right)^3 
\int_{0}^{\infty} dv_1\, \partial_{v_1} \Phi_{\text{LTW}}(v_1) \Psi_2(v_1)
\int_{0}^{v_1} dv_2\, \partial_{v_2} \Phi_{\text{LTW}}(v_2) \Psi_2(v_2) \nonumber \\
& \phantom{ \left( -\frac{ig}{\hbar} \right)^3  \quad} \times \int_{0}^{h(v_2)} dv_3\, \partial_{v_3} \Phi_{\text{LTW}}(v_3) \Psi_1(v_3) \label{LLL221}
\end{align}

\begin{align}
LLL_{222} 
=& \left( -\frac{ig}{\hbar} \right)^3 
\int_{0}^{\infty} dv_1\, \partial_{v_1} \Phi_{\text{LTW}}(v_1) \Psi_2(v_1)
\int_{0}^{v_1} dv_2\, \partial_{v_2} \Phi_{\text{LTW}}(v_2) \Psi_2(v_2) \nonumber \\
& \phantom{ \left( -\frac{ig}{\hbar} \right)^3  \quad} \times \int_{0}^{v_2} dv_3\, \partial_{v_3} \Phi_{\text{LTW}}(v_3) \Psi_2(v_3) \label{LLL222}
\end{align}

\bibliographystyle{jhep}
\bibliography{MainUnruhRef}
\end{document}